\documentclass[letterpaper,12pt,peerreviewca,draftcls]{IEEEtran}
\usepackage{csm16}
\usepackage[margin=1in]{geometry}
\usepackage[nolists,nomarkers,tablesfirst]{endfloat} 
\usepackage{amsmath,amsfonts,amssymb,mathtools} 

\usepackage[labelformat=simple]{subcaption}

\usepackage{algorithmic}
\usepackage{url}
\usepackage{graphicx,xcolor}
\usepackage{verbatim}
\makeatletter
\newcommand{\verbatimfont}[1]{\def\verbatim@font{#1}}%
\makeatother
\verbatimfont{\ttfamily\small}

\newcommand{\bi}{\begin{itemize}}\newcommand{\ei}{\end{itemize}}
\newcommand{\be}{\begin{equation}}\newcommand{\ee}{\end{equation}}
\newcommand{\bee}{\begin{enumerate}}\newcommand{\eee}{\end{enumerate}}
\newcommand{\bea}{\begin{eqnarray}}\newcommand{\eea}{\end{eqnarray}}
\newcommand{\beas}{\begin{eqnarray*}}\newcommand{\eeas}{\end{eqnarray*}}
\newcommand{\bc}{\begin{center}}\newcommand{\ec}{\end{center}}
\usepackage[left,pagewise]{lineno} 
\usepackage[english]{babel} 
\usepackage{blindtext}

\newcommand{\Expect}{\mathbb{E}}
\newcommand{\E}{\mathbb{E}}
\newcommand{\ud}{\,\mathrm{d}}

\newcommand{\Xbar}{\bar{X}}

\newcommand{\calV}{\mathcal{V}}

\newcommand{\K}{{\sf K}}

\newcommand{\clZ}{\mathcal{Z}}
\newcommand{\NN}{\mathcal{N}}
\newcommand{\newP}[1]{\medskip\noindent{\bf{#1}}}

\newcommand{\KN}{{\sf K}^{(N)}}

\newcommand{\mN}{m^{(N)}}
\newcommand{\SigN}{\Sigma^{(N)}}
\def\Re{\mathbb{R}}
\newcommand{\half}{\frac{1}{2}}
\newcommand{\Teps}{T_\epsilon}
\newcommand{\TepsN}{T_\epsilon^{(N)}}
\newcommand{\Ten}{{\sf T}}
\newcommand{\hvec}{{\sf h}}
\newcounter{rmnum}

\newtheorem{proposition}{Proposition}

\newenvironment{romannum}{\begin{list}{{\upshape (\roman{rmnum})}}{\usecounter{rmnum}
			\setlength{\leftmargin}{12pt}
			\setlength{\rightmargin}{8pt}
			\setlength{\itemsep}{2pt}
			\setlength{\itemindent}{-1pt}
	}}{\end{list}}

\newcounter{anum}

\title{Optimal Transportation Methods \\ in Nonlinear Filtering\\
	\Large The feedback particle filter}
\author{Amirhossein Taghvaei and Prashant G. Mehta\\ \today}

\newif\ifPDF \ifx\pdfoutput\undefined\PDFfalse \else\ifnum\pdfoutput > 0\PDFtrue \else\PDFfalse \fi \fi
\ifPDF 
\usepackage[pdftex, plainpages = false, colorlinks=true, linkcolor=black, citecolor = green!50!blue, urlcolor = blue, filecolor=black, pagebackref=false, hypertexnames=false,  pdfpagelabels ]{hyperref}
\fi

\begin{document}
	\maketitle
	\CSMsetup

{\em How Data Became One of the Most Powerful Tools to Fight an
  Epidemic} is a question that a recent (Jun 10 2020) NYT magazine
article poses in its title and addresses in its content.  Indeed, the
spread of COVID-19 involves dynamically evolving hidden data (e.g.,
number of infected, number of asymptomatic etc.) that must be
deduced from noisy and partially observed data (e.g., number of daily
deaths, number of daily hospitalized, number of daily tested positive etc.).   
The underlying mathematics for posing and solving this and 
several other partially observed dynamic problems is familiar to most
control theorists.

A mathematical abstraction of these types of problems commonly
involves definition of two stochastic processes $(X,Z)$ where in
continuous-time settings $X=\{X_t\in\mathbb{S}:t\geq 0\}$ is the
hidden signal process and $Z=\{Z_t\in\mathbb{O}:t\geq 0\}$ is the
observed or measured process.  For the sake of exposition, the
state-space $\mathbb{S}$ and the observation-space $\mathbb{O}$ are
assumed to be the Euclidean spaces, and the two processes are modeled as
solution of a stochastic differential equation (SDE)
\begin{align}
\ud X_t  & = a (X_t) \ud t + \sigma(X_t) \ud B_t,\quad X_0 \sim
  \text{(prior)},     \label{eq:Xt}\\
\ud Z_t & = h(X_t) \ud t + \ud W_t, \quad Z_0=0, \label{eq:Zt}
\end{align}
where $a(\cdot), \sigma(\cdot), h(\cdot)$ are given smooth
functions of their arguments, the signal (or process) noise
$B=\{B_t:t\geq 0\}$ 
and the measurement (or observation) noise $W=\{W_t:t\geq
0\}$ are assumed to be independent Wiener processes (w.p).  For
example, in the models of disease spread, $Z_t$ may indicate the number
(cumulative) of tested positive up to time $t$. In this case,
$Z_{t_2}-Z_{t_1}$ is the number (increment) of tested positive during
the time interval $[t_1,t_2]$, and ``$\ud Z_t$'' can be thought of as
the infinitesimal increment over the infinitesimal time ``$\ud t$''.

Given models for the stochastic processes $(X,Z)$, the mathematical
problem of stochastic filtering is to estimate the
conditional distribution of the state $X_t$ given observations up to
time $t$.  The conditional distribution ${\sf P}(X_t|\clZ_t)$ is
referred to as the posterior distribution where $\clZ_t$ is the
time-history (filtration) of observations up to time $t$.

Owing the widespread importance of this problem, there are a host of
solution approaches under different modeling assumptions.  The most
classical of these approaches is the method of least squares.  The
method was invented at the turn of the 19th century but remains
popular to date, for instance, in identification of the (static) model
parameters (see~\cite{bastos2020modeling} for an application of these
methods to parameter estimation in disease modeling).  For the dynamic case, when the models are linear (i.e., $a(x) = A x$, $\sigma(x) = \sigma$ and $h(x) = H x$) and the distributions (of the noise processes and the prior) are Gaussian,
Kalman and Bucy derived a recursive algorithm~\cite{kalman-bucy} known
as the Kalman-Bucy filter
\[
\ud \hat{X}_t = \underbrace{A \hat{X}_t \ud t}_{\text{dynamics}}   \; +\;  \underbrace{K_t (\ud Z_t - H \hat{X}_t \ud t)}_{\text{control}},
\]
where $\hat{X}_t := {\sf E}(X_t|\clZ_t)$ is the conditional mean and $K_t$ is the Kalman gain.  Each of the two terms on the right-hand side have an intuitive explanation.  The first term accounts for the effect of the dynamics due to the signal model.  The second term implements the effect of conditioning because of the most recent observation (increment) ``$\ud Z_t$''.  The second term is referred to as the correction or the Bayes' update step of the Kalman filter.  

It is remarkable that the Bayes' update formula in the Kalman filter 
takes the form of a feedback control law where   
\[
[\text{control}] = [\text{gain}] \cdot [\text{error}],
\]
and 
\begin{align*}
[\text{error}]  = [\text{Observation}] -  \,[\text{prediction}].
\end{align*}
Note  that ``$H \hat{X}_t\ud t$" is the filter prediction of the new observation ``$\ud Z_t$''.  The formula is so simple that it can and should be part of any introductory undergraduate controls class -- as an example of
proportional gain feedback control law!  Of course, this simple formula 
has had an enormous impact in many applications such as target tracking and surveillance,
  air traffic management,  weather surveillance, ground mapping,
  geophysical surveys,  remote sensing, autonomous navigation and
  robotics.

The Kalman filter has many extensions, for example, to problems
involving additional uncertainties in the signal and the observation
models. The resulting algorithms are referred to as the interacting
multiple model (IMM) filter~\cite{Blom_cdc12} and the probabilistic
data association (PDA) filter~\cite{Bar-Shalom_IEEE_CSM},
respectively. In the PDA filter, the Kalman gain is allowed to vary
based on an estimate of the instantaneous uncertainty in the
observations. In the IMM filter, multiple Kalman filters are run in
parallel and their outputs combined to form an estimate.

Arguably, the structural aspects of the Kalman filter have been as
  important as the algorithm itself in design, integration, testing
  and operation of the overall system.  As a simple illustration of
  this, consider for example the Kalman filter gain.  The gain is known to
  scale proportionally to the signal-to-noise ratio (SNR) of the
  observations.  In practice, the gain is often tuned or adapted in an
  online manner to trade-off performance for robustness.  Without such
  structural features, it is a challenge to create scalable
  cost-effective robust solutions.

A limitation of the Kalman filter is that it gives exact solution only in the linear
Gaussian settings.  Beginning in early 1990s, spurred in part by
computational advances, simulation-based Monte-Carlo (MC) algorithms
became popular for the purposes of numerically approximating the
posterior distribution in more general settings.  These class of
algorithms, referred to as particle filters, approximate the posterior
distribution using a population of $N$ particles $\{X_t^i:t\geq 0,
1\leq i \leq N\}$. One can interpret each of the particles as
independent samples drawn from the posterior. Alternatively, one can
interpret the empirical distribution of the population as
approximating the posterior distribution.

Like the 
Kalman filter, the particle filter too is a recursive algorithm.  The
signal model is used to simulate the effect of the dynamics.  The
Bayesian update step is implemented using techniques such as
importance sampling and resampling.  Although these techniques are
easily described, they bear little resemblance to the feedback control
structure of the Kalman filter.

\begin{figure}[t]
	\centering
	\begin{tabular}{cc}
		\begin{subfigure}{.5\textwidth}
			\includegraphics[width = 0.95\hsize]{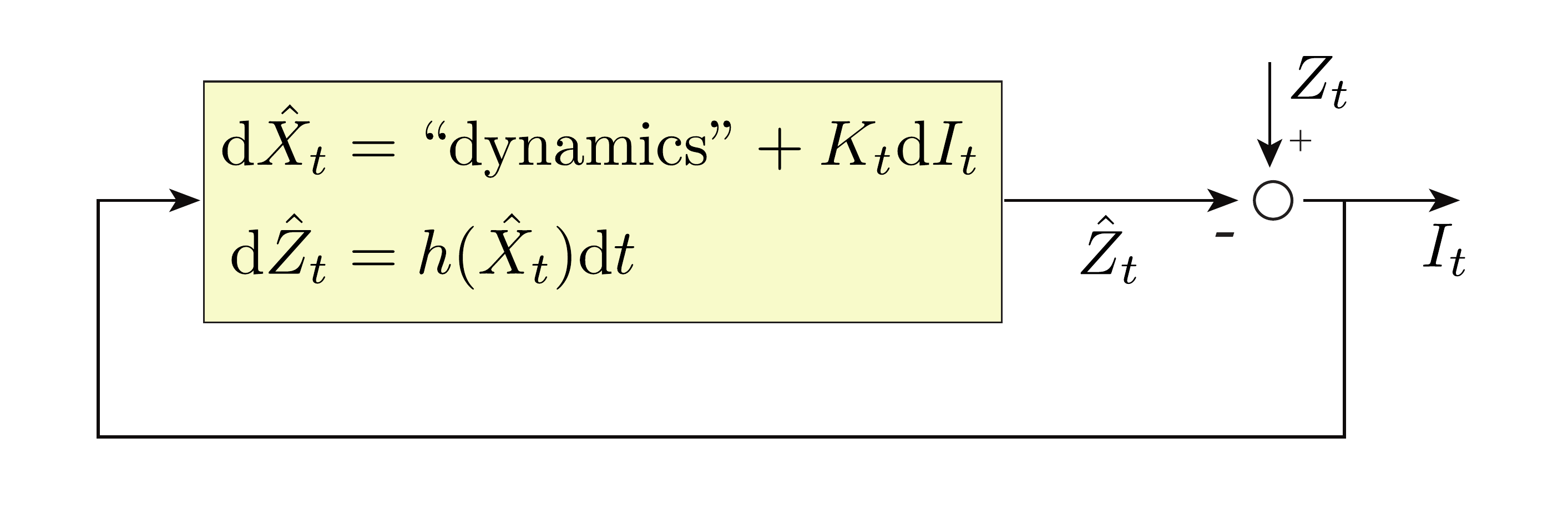}
				\subcaption[aaa]{Kalman filter}
			\label{fig:KF-diagram} 
		\end{subfigure} &
		\begin{subfigure}{.5\textwidth}
			\includegraphics[width = 0.95\hsize]{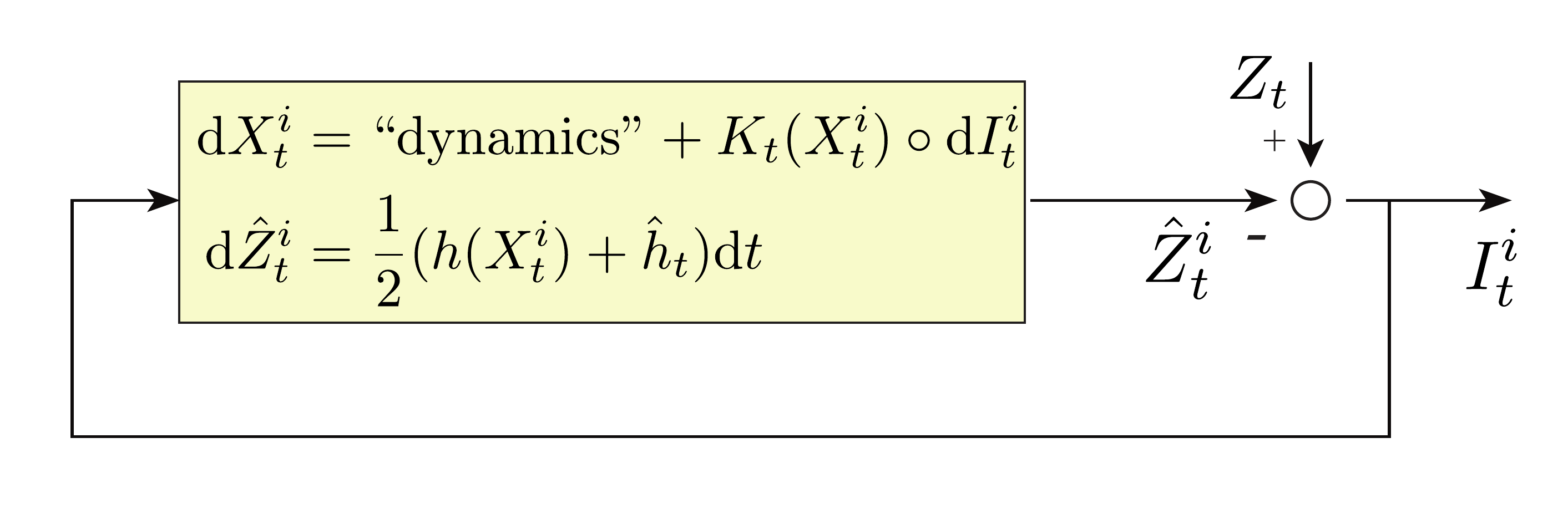}
			\subcaption{Feeedback particle filter}
			\label{fig:FPF-diagram}
		\end{subfigure}
	\end{tabular}
	\caption{Feedback control structure of the Kalman filer and
          the feedback particle filter (FPF). $\hat{X}_t$ in the KF is the estimate (conditional mean) of the hidden state. $X_t^i$ in the FPF is a sample from the posterior (conditional distribution) of the hidden state.  In either algorithms, the Bayesian update is implemented via a gain times error control law. } 
	\label{fig:filter-diagram}
\end{figure}

The focus of the present paper is on the feedback particle filter
(FPF) algorithm.  FPF represents an exact solution of the nonlinear
non-Gaussian filtering problem~\eqref{eq:Xt}-\eqref{eq:Zt} where the
state-space $\mathbb{S}$ can in general be a Riemannian manifold.  (In
applications, Euclidean spaces and matrix Lie groups are most
common.) The
distinguishing feature of the FPF is that the Bayesian update step is
implemented via a feedback control law of the form
\[
[\text{control}] = [\text{gain}] \cdot [\text{error}],
\]
where
\begin{align*}
[\text{error}]  = [\text{Observation}] -  \left( \frac{1}{2}\,
  [\text{Part. predict.}] + \frac{1}{2} \,[\text{Pop. predict.}] \right).
\end{align*}
The terms [Part. predict.] and [Pop. predict.] refer to  
the prediction -- regarding the next [Observation] ``$\ud Z_t$'' -- as made by the particle and by the population, respectively (see the diagram~\ref{fig:filter-diagram}).  Because the control for each particle depends also on the population (and thus the empirical distribution), this is an example of a mean-field type control law; cf.,~\cite{bensoussan2013mean,carmona2018probabilistic}.  
It turns out that for the linear Gaussian problem in the Euclidean state-space, the $[\text{gain}]$ of
the FPF is exactly the Kalman gain.  In
non-Gaussian settings, the gain solves  a certain linear partial
differential equation (PDE) known as the weighted Poisson equation.
The exact formula for the FPF control and gain appears in the main
body of the paper.

At the turn of the decade beginning 2010, the FPF algorithm was
introduced by our research group at the University of
Illinois~\cite{yanmehmey11,yanmehmey11b,yang2013feedback}.  The algorithm can be viewed as a modern extension
to the Kalman filter, a viewpoint stressed in a prior review
paper~\cite{TaghvaeiASME2017}.  Like the Kalman filter, the FPF is easily
  extended to handle additional uncertainties in signal and
  measurement model: 
  These extensions, namely, the joint probabilistic data association
  (JPDA)-FPF and the interacting multiple model (IMM) FPF appear
  in our prior works~\cite{yang2012joint,yang2013interacting}.  

From a historical
perspective, FPF is part of a broader
class of exact and approximate interacting particle algorithms, in particular,          
the ensemble Kalman filter (EnKF) which is widely used for data
assimilation in weather prediction and other types of geo-physical
applications~\cite{evensen1994sequential}.  Closely related to and pre-dating our work on FPF, the first
interacting particle representation of the continuous-time nonlinear
filtering problem~\eqref{eq:Xt}-\eqref{eq:Zt} appeared in~\cite{crisan10}.   In linear Gaussian settings,
the update formula for FPF is known as the square
root form of the EnKF~\cite{reich11,Reich-ensemble}.

The objective for this paper is to situate the development of FPF and
related controlled interacting particle system algorithms (e.g., EnKF)
within the framework of optimal transportation theory.  The key notion
is that of ``coupling" between two distributions -- prior and
posterior in the Bayesian settings of this paper.  Optimal
transportation theory is then applied to design the optimal coupling.
Of course, in practice, this requires a solution of certain PDEs --
such as the Poisson equation that arise in the FPF algorithm.  The
coupling viewpoint has several advantages which are  described in the
main body of this paper.

 \section{The coupling viewpoint}
 
The heart of any simulation-based recursive particle filter algorithm is the Bayes' update formula
\[
[\text{posterior}] \propto [\text{likelihood}] \cdot [\text{prior}].
\]
The notation $p_0(\cdot)$, $p_1(\cdot)$, and $\ell(\cdot)$ is used to
denote the prior, posterior, and likelihood distributions
respectively.  The expressions for these in the linear Gaussian
example appear in ``optimal coupling for  Gaussian distributions".

In any particle filter algorithm, one also needs to simulate the
effect of dynamics.  This step is straightforward using the signal
model directly.  Therefore, we focus here only on the update formula.

In simulation settings, one of the challenges is that an analytic
expression of the prior distribution is not available.  Instead, the
prior is approximated in terms of $N$ independent samples
$\{X^i_0:1\leq i\leq N\}$:
	\begin{equation*}
	p_0(x) \approx\frac{1}{N}\sum_{i=1}^N \delta_{X^i_0}(x),
	\end{equation*}    
	where $\delta_z$ is the Dirac distribution at $z\in\mathbb{S}$.
	The expression on the right-hand side is referred to as the
        empirical distribution of the population.  Alternatively, one
        can think of $X_0^i$ as independent samples drawn from the
        prior.  

\begin{figure}
	\centering
	\includegraphics[width=0.8\hsize]{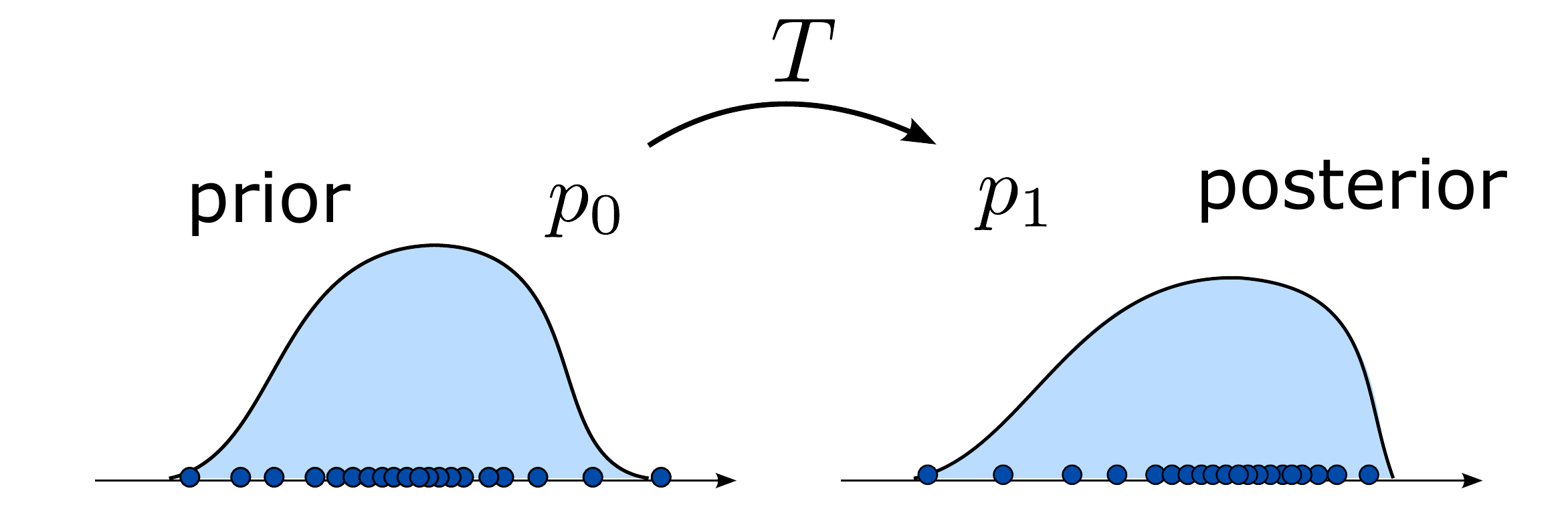}
	\caption{Coupling viewpoint of the filtering problem. The task of a particle filter is to convert a sample of $N$ particles from the prior distribution to a sample of $N$ particles from the posterior distribution. This task is viewed as finding a coupling between the prior and the posterior distributions. }
	\label{fig:transport}
\end{figure}

In a numerical implementation of the update formula, the problem is to convert the sample of $N$ particles  $\{X^i_0:1\leq i\leq N\}$ from the prior distribution $p_0(\cdot)$ to a sample of $N$ particles  $\{X^i_1:1\leq i\leq N\}$ from the posterior distribution $p_1(\cdot)$.  The algorithmic problem is depicted in Figure~\ref{fig:transport} and expressed as follows
		 \begin{align*}
	\text{Input:}&\quad \text{samples}~\{X^i_0:1\leq i\leq N\} \sim p_0,\quad \text{likelihood function}~\ell,\\	
	\text{Output:}&\quad \text{samples}~\{X^i_1:1\leq i\leq N\} \sim p_1.
	\end{align*}

	The task of converting samples from one distribution
        $p_0(\cdot)$ to samples from another distribution $p_1(\cdot)$
        is viewed here as the problem of finding a coupling
        $\pi(\cdot,\cdot)$;
        cf.,~\cite{reich2015probabilistic,reich2019data,reich13}.  By
        definition, a coupling is any joint probability distribution
        that satisfies the marginal constraints $\int \pi(x,x') \ud x'
        = p_{1}(x)$ and $\int \pi(x,x') \ud x = p_{0}(x')$.   It is
        convenient to express $\pi(x,x') = \mathcal{T}(x|x')\,p_0(x')$ where $\mathcal{T}(\cdot|\cdot)$ is referred to as the transition kernel. 
	Once the coupling is at hand, new samples are generated by using the transition kernel. 
	
	Given this viewpoint, the MC algorithm simulates the following stochastic update law for the system of particles:
	\begin{equation}\label{eq:intro-PF-init}
	X^i_{1} \sim \mathcal{T}(\cdot |X^i_0).
	\end{equation}
	This means that a new sample $X^i_{1}$ is generated by sampling from the distribution $\mathcal{T}(\cdot |X^i_0)$.  
	The sampling algorithm~\eqref{eq:intro-PF-init} ensures that if the probability
        distribution of $X^i_0$ is $p_0$ then the probability distribution of $X^i_1$ is $p_1$.
	The associated algorithmic task is expressed as
	\begin{align*}
	\text{Input:}&\quad \text{samples}~\{X^i_0:1\leq i\leq N\} \sim p_0,\quad \text{likelihood function}~\ell,\\	
	\text{Output:}&\quad \text{coupling between $p_0$ and $p_{1}$.}
	\end{align*}

\subsection{SIR particle filter}
	There are infinitely many couplings between two
        distributions. The simplest possible choice is an independent
        coupling where $ \pi(x,x') = p_{1}(x) p_0(x')$. For the
        independent coupling, the transition kernel
        $\mathcal{T}(x|x')= p_{1}(x)$.  Sequential importance
        resampling (SIR) particle filters~\cite{gordon93,doucet09}
        numerically implement the independent coupling in two steps:
	\begin{romannum}
		\item[Step 1:]    	 A weighted distribution of
                  the particles is first formed according to
                  $\sum_{i=1}^N w_i\delta_{X^i_0}$ where the weights
                  $w_i=\frac{l(X^i_0)}{\sum_{j=1}^N l(X^j_0)}$. The
                  weighted distribution is an approximation of the
                  posterior distribution  $p_{1}$.  This step is
                  called importance sampling. 
		\item[Step 2:]  	 Next $N$ particles are
                  independently sampled from the weighted
                  distribution: $X^i_{1} \sim \sum_{i=1}^N
                  w_i\delta_{X^i_{0}}$, by sampling from a multinomial
                  distribution with parameter vector
                  $(N,\{w_i\}_{i=1}^N)$. This step is called
                  resampling.
	\end{romannum}

 Theoretically, it is shown that the empirical approximation with
 particles becomes exact in the limit as $N \to \infty$ with error
 rate $O(N^{-1/2})$~\cite{del2001stability,cappe2009inference}.
 However, both empirically and theoretically, it was discovered that
 particle filters can suffer from large simulation variance that
 progressively gets worse as the dimension of the problem
 increases~\cite{sr:bengtsson08,beskos2014error,rebeschini2015can}. To
 maintain the same mean-squared error, a particle filter is known to
 require a number of particles that scales exponentially with the
 dimension. This issue is referred to as the {\em curse of
   dimensionality} (CoD).

The issues with the stochastic independent coupling in SIR filters has
motivated investigation of other forms of coupling that are also optimal
in some sense~\cite{delmoralbook,bain2009}.  In the simulation
literature, this is referred to as the design of proposal
distributions.  
	
\subsection{Optimal transport coupling}

Optimal transportation theory provides a principled approach to
identify a coupling.  Given two distributions $p_0$ and $p_1$,
the optimal transportation problem is
\begin{equation}
\min_{\pi  \in \Pi_{p_0,p_1}}~\Expect_{(X_0,X_1)\sim \pi}[|X_0-X_1|^2],\label{eq:Kantorovich-p}
\end{equation}
where $\Pi_{p_0,p_1}$ is the set of all couplings with
marginals fixed to $p_0$ and $p_1$.  The optimal cost is referred to as $L^2$-Wasserstein distance between $p_0$ and $p_1$~\cite{villani2003topics}. 

The optimization problem~\eqref{eq:Kantorovich-p} is known as the
Kantorovich formulation of the optimal transportation problem. By a
famous result due to Brenier, if the two distributions admit density
with respect to Lebesgue measure, then the optimal coupling is unique
and deterministic of the form  $\pi(x,x') = \delta_{x=\nabla
  \Phi(x')}p_0(x')$ where $\Phi$ is a convex function~\cite[Thm.
2.12]{villani2003topics}. The function $\Phi$ is obtained by solving
the Monge-Amp\`ere PDE~\cite{evans}. A numerical solution to the
Monge-Amp\`ere PDE based only on the samples is a challenging
problem. In the following, we discuss the special setting where the
the prior and posterior distributions are Gaussian.

  \subsection{Optimal coupling for Gaussian distributions}
  
In Gaussian settings, the prior and the likelihood are both Gaussian distributions:
\[
p_0(x)\propto \exp(-\frac{1}{2}(x-m_0)^\top \Sigma_0^{-1}(x-m_0)),\quad 
l(x) \propto \exp(-\frac{|y-Hx|^2}{2}).
\]
In this case, a simple completion of square helps show that the posterior is also a Gaussian distribution
\[
p_1(x)\propto \exp(-\frac{1}{2}(x-m_1)^\top \Sigma_1^{-1}(x-m_1)).
\]
This yields the following update formula for the mean and variance:
	\begin{align*}
	m_{1} &= m_0 + \K(Y - Hm_0),\quad 
	\Sigma_{1} =  \Sigma_0 - \K H \Sigma_0,
	\end{align*}
where $\K = \Sigma_0 H^\top (H\Sigma_0H^\top + I)^{-1}$ is the Kalman
gain.  This is in fact the update formula for the discrete-time Kalman filter.

The coupling design problem is to couple the Gaussian prior $p_0$ and
the Gaussian posterior $p_1$.  The optimal coupling, 
solution to the optimal transportation
problem~\eqref{eq:Kantorovich-p}, between two Gaussian distributions is explicitly known and is an affine map of the form
\begin{equation}\label{eq:OT-map-Gaussian}
T(x) = F(x-m_0) + m_{1},  
\end{equation}
where $F$ is the (unique such) symmetric matrix solution to the matrix
equation $F\Sigma_0F = \Sigma_{1}$. 
The explicit form of the solution
is \[F=\Sigma_0^{-\frac{1}{2}}\left(\Sigma_0^{\frac{1}{2}}\Sigma_1
    \Sigma_0^{\frac{1}{2}}\right)^{\frac{1}{2}}\Sigma_0^{-\frac{1}{2}}.\]
 Note that if $X \sim  \NN(m_0,\Sigma_0)$ then $T(X) \sim
\NN(m_{1},\Sigma_{1})$ because: (i) the mean $\Expect[T(X)] =
m_{k+1}$; (ii)  the variance $\Expect[(T(X)-m_{1})(T(X)-m_{1})^\top] =
F \Sigma_0 F = \Sigma_{1}$; (iii) and an affine transformation of a
Gaussian random variable is again Gaussian.

The optimal transport map~\eqref{eq:OT-map-Gaussian} yields the
following algorithm for sampling
\begin{equation*}
X_1^i  = T({X}_0^i) = F(X_0^i -m_0) + m_{1}.
\end{equation*}
Given $X_0^i\sim p_0$ we have $X_1^i\sim p_1$.  

The optimal coupling depends upon the statistics of both 
the prior and the posterior distributions.  In simulation-based
setting, one only has a population of particles $\{X_0^i:1\leq i \leq
N\}$.  So, the transport map needs also to be approximated from the
particles.  One such approximation is as follows:
\begin{equation}\label{eq:Xi-OT-discrete-particles}
\text{particle update:}\quad X^i_{1} = F^{(N)}(X^i_0-\mN_0) + \mN_0 + \KN(Y-H\mN_0),
\end{equation}
where $\mN_0 = \frac{1}{N}\sum_{i=1}^NX^i_0$ is an empirical approximation of the mean $m_0$, $\SigN_0=\frac{1}{N-1}\sum_{i=1}^N (X^i_0-\mN_0)(X^i_0-\mN_0)^\top$ is an empirical approximation of the variance, $F^{(N)}$ is the unique symmetric matrix solution to the matrix equation $F^{(N)}\SigN_0F^{(N)} =\SigN_0 - \KN H \SigN_0$, and $\KN = \SigN_0H^\top(H\SigN_0H^\top +  I)^{-1}$.

The update formula~\eqref{eq:Xi-OT-discrete-particles} is compared to
the discrete-time EnKF update:
\begin{equation}\label{eq:Xi-EnKF-discrete-particles}
\text{particle update (EnKF):}\quad X^i_{1} =X^i_0 + \KN(Y-HX^i_0+W^i),
\end{equation}
where $\{W^i:1\leq i\leq N\}$ are independent copies of the
observation noise.  The EnKF update is an example of a
stochastic coupling in contrast to the deterministic 
optimal coupling~\eqref{eq:Xi-OT-discrete-particles}.
The EnKF update does not require solving for $F^{(N)}$. This makes it
simpler to implement numerically.  However, the presence of noise
$W^i$ in
the update law introduces an additional source of error in any finite-$N$
implementation~\cite{AmirACC2016}.  

Besides~\eqref{eq:Xi-EnKF-discrete-particles}, there are several
other forms of the EnKF update.  One particular update -- that has been
crucial in successful application of EnKF in geosciences -- is the
ensemble square-root Kalman filter
(EnSRKF)~\cite{whitaker2002ensemble}.  This update is an example of a
deterministic coupling that also avoids the need to compute
$F^{(N)}$~\cite[Sec. 7.1]{reich2015probabilistic}.  In the sequel,
the continuous-time  version of the EnSRKF is shown to arise as a special case of the FPF.          


The systems~\eqref{eq:Xi-OT-discrete-particles} and~\eqref{eq:Xi-EnKF-discrete-particles} are both examples of interacting
  particle system. The interaction arises because of  the terms
  involving the empirical quantities $\mN_0$ and $\SigN_0$. In the
  limit as $N\to \infty$, these converge to their respective
  statistics and the particles become independent of each other~\cite{mandel2011convergence,delmoral2016stability,delmoral2017stability,jana2016stability,Taghvaei2019AnOT}. This
  phenomenon is referred to as the propagation of
  chaos~\cite{sznitman1991,rachev1998}. The ($N=\infty$) limit is
  referred to as the mean-field limit and
  for~\eqref{eq:Xi-OT-discrete-particles} it is identified with a single equation
\begin{equation}\label{eq:Xi-OT-discrete-mf}
   \text{mean-field update:}\quad  \Xbar_1 = F(\Xbar_0-m_0) + m_0 + \K(Y-Hm_0).
\end{equation}

As a practical matter, one first designs a
coupling~\eqref{eq:OT-map-Gaussian} which is used to define the
mean-field process~\eqref{eq:Xi-OT-discrete-mf}.  Subsequently, the
mean-field terms are approximated empirically to define a particle
system~\eqref{eq:Xi-OT-discrete-particles}.  A finite-$N$ implementation of the particle system yields a practical algorithm to solve the filtering task.

A summary of this subsection is presented in the sidebar ``Summary of the linear Gaussian example".
An excellent exposition of the coupling approach to discrete-time
filtering appears in the 
monograph~\cite{reich2015probabilistic} and the tutorial style review
article~\cite{reich2019data}.  Other
examples of couplings include the approximation of
the Rosenblatt transport
map~\cite{MarzoukBayesian,mesa2019distributed} and Gibbs flow~\cite{heng2015gibbs}.

\section{Feedback particle filter}

The coupling viewpoint is next employed to introduce and describe the
feedback particle filter (FPF) algorithm.  FPF is a MC solution to the
continuous-time nonlinear filtering
problem~\eqref{eq:Xt}-\eqref{eq:Zt}.

A construction of FPF proceeds in the following two steps:
\begin{romannum}
	\item[Step 1:] Construct a stochastic process
          $\bar{X}=\{\bar{X}_t \in \mathbb{S}:t\geq0\}$ such that the 
          conditional distribution (given $\clZ_t$) of $\bar{X}_t$ is equal to the posterior distribution of $X_t$; 
	\item[Step 2:] Simulate $N$ stochastic processes
          $\{X^i_t \in \mathbb{S}:t\geq 0,1\leq i\leq N\}$ to empirically
          approximate the conditional distribution of $\bar{X}_t$.
\end{romannum}
\begin{equation*}
\underbrace{\Expect[f(X_t)|\clZ_t]\overset{\text{Step
			1}}{=}\Expect[f(\bar{X}_t)|\clZ_t]}_{\text{exactness condition}}
\overset{\text{Step 2}}{\approx} \frac{1}{N}\sum_{i=1}^N f(X^i_t)\label{eq:exactness}.
\end{equation*}
for all bounded functions $f$.
The process $\bar{X}$ is referred to as mean-field process and the
$N$ processes are referred to as particles.  The
construction ensures that the filter is {\em exact} in the mean-field
($N=\infty$) limit.    

\newP{Mean-field process:}
The mean-field process $\bar{X}$ is modeled as a solution of a controlled SDE:
\begin{equation}
\ud \Xbar_t = \underbrace{a(\bar{X}_t) \ud t + \sigma(X_t) \ud \bar{B}_t}_{\text{dynamics}} + \underbrace{u_t(\Xbar_t)\ud t + \K_t(\Xbar_t) \ud Z_t}_{\text{control}},\quad \Xbar_0 \stackrel{\text{d}}{=} X_0,
\label{eq:u-k}
\end{equation}
where $\bar{B}=\{\bar{B}_t:t\geq 0\}$ is a w.p. independent of $\bar{X}_0$.  
The first two terms in the SDE~\eqref{eq:u-k} are a copy of the
dynamics~\eqref{eq:Xt}. The other two terms are control laws
(transition kernels) 
that need to be designed to implement the filtering update step: The mathematical control objective is to
design $\{u_t(\cdot):t\geq 0 \}$ and $\{\K_t(\cdot):t\geq 0 \}$ such that
the conditional distribution (given $\clZ_t$) of $\bar{X}_t$ is equal
to the posterior distribution of $X_t$.

The control is regarded as implementing the transition kernel of a coupling.  
As in the simpler discrete-time setting of the preceding section, there are infinitely many couplings and associated transition kernels that satisfy the exactness criteria. This is not surprising: The exactness condition specifies only the
marginal distribution of $\Xbar_t$ at times $t\geq 0$. This is clearly
not enough to uniquely identify a stochastic process, for instance,
the joint distributions at two time instants are not specified.

The procedure from the preceding section is suitably adapted to design
the optimal coupling.  The optimality criterion is the Kantorovich
form~\eqref{eq:Kantorovich-p} of the optimal transportation problem.   
The details appear in the sidebar ``Optimal
transport construction of stochastic processes" where it is shown that
the optimal $\K_t$ is of the gradient form as follows:
\begin{align*}
    \K_t(x) &= \nabla \phi_t(x),\quad \text{where $\phi_t$ solves the PDE}\quad  -\nabla \cdot(p_t \nabla\phi_t) = p_t(h-\hat{h}_t),
\end{align*}
where $\nabla$ is the gradient operator, $\nabla \cdot $ is the
divergence operator, $\hat{h}_t := \E[h(\bar{X}_t)|\mathcal{Z}_t]
=\int h(x) p_t(x) \ud x$, and $p_t$ is the conditional density of
$\bar{X}_t$.

The optimal solution for $u_t$ is as follows:
\begin{equation*}
   u_t(x) = -\frac{1}{2}(h(x) +  \hat{h}_t)\K_t(x) +\frac{1}{2}\K_t \nabla \K_t(x)  + \xi_t(x), 
\end{equation*}
where $\xi_t$ is the (unique such) divergence-free vector field (i.e.,
$\nabla
\cdot(p_t \xi_t) = 0$) such that $u_t$ is of a gradient form.
An intuitive explanation of the three terms is as follows: The first
term is the gain times the average at the `particle' prediction $h(x)$ and the
population prediction $\hat{h}_t$.  Together with the stochastic
term $\K_t\,\ud Z_t$, the first term yields the gain times error structure of the
FPF.  The second term is the so-called Wong-Zakai correction term from
which it follows that the gain times error formula is 
expressed in its Stratanovich form.  (The geometric significance of
the Stratanovich form is described after the FPF formula has been formally presented.) 
The significance of the third term is the subject of the next
paragraph.

It turns out that the divergence-free choices of $\xi_t$ parametrize a
manifold of couplings all of which yield the same (exact)
posterior. Therefore, the choice of $\xi_t$ affects {\em only} the
(Wasserstein) optimality but not the exactness property of the filter.
In FPF, we simply take $\xi_t\equiv 0$ for all $t\geq 0$.  Although such a solution is optimal
only in the scalar (one-dimensional) settings, it avoids the need to solve an additional
PDE to compute the optimal $\xi_t$.  The resulting algorithm is referred
to as the feedback particle filter.  It is described next.  

\newP{Feedback particle filter:}
The mean-field process $\bar{X}$ evolves according to the SDE:
\begin{equation}
\ud \bar{X}_t = 
\underbrace{a(\bar{X}_t) \ud t + \sigma(X_t) \ud \bar{B}_t}_{\text{dynamics}} + \underbrace{\K_t(\bar{X}_t) \circ (\ud Z_t -
	\frac{h(\bar{X}_t) + \hat{h}_t}{2}\ud t)}_{\text{Bayes' update: feedback control law}},
\quad \bar{X}_0\sim p_0,
\label{eq:FPF-mean-field}
\end{equation}  
where the symbol $\circ$ is used to denote the fact that the SDE is
expressed in its Stratanovich
form~\cite[Sec. 3.3]{oksendal2013}.
 The It\^{o}
  form of the FPF includes the standard Wong-Zakai correction term
  that arises on account of the dependence of the gain $\K_t(\cdot)$
  on the state $X_t$~\cite[Eq. 2]{yang2016}. Because the gain also depends upon
  the density, the interpretation of the Stratonovich form in the general case is more
  involved as discussed at length in~\cite{pathiraja2020mckean}.
   
The gain function is $\K_t(x):= \nabla \phi_t(x)$ where $\phi_t$ is the solution of the Poisson
equation:
\begin{equation}
\label{eq:Poisson-intro}
\begin{aligned}
\text{Poisson equation:}&\quad - \frac{1}{p_t(x)}\nabla \cdot (p_t(x)
\nabla \phi_t(x) ) = (h(x)-\hat{h}_t),\quad\forall \;x\in\Re^d.
\end{aligned}
\end{equation}
The operator on the left-hand side of~\eqref{eq:Poisson-intro} is the
{\em probability-weighted Laplacian}.  It is denoted as
$\Delta_{\rho}:=\frac{1}{\rho} \nabla \cdot (\rho \nabla )$
where at time $t$ the probability density $\rho$ is the conditional
density $p_t$.  The equation~\eqref{eq:Poisson-intro} is referred to
as the Poisson equation of nonlinear filtering~\cite{laugesen15}.

The Stratanovich form of the update formula provides for an intrinsic
(i.e., coordinate independent) description of the filter.   It was shown that the FPF is an exact filter not only
in the Euclidean settings but also when the state-space
$\mathbb{S}$ is a Riemannian
manifold, e.g., a matrix Lie group~\cite{zhang2017feedback}.  
For a
manifold with boundaries, the Poisson equation is supplemented with
the Neumann boundary conditions.

\newP{Particles:} A finite-$N$ algorithm is obtained by empirically
approximating the mean-field control law. The particles $\{X^i_t:t\geq
0 , 1\leq i\leq N\}$ evolve according to:
\begin{equation*}
\ud X^i_t = a(X^i_t) \ud t + \sigma(X^i_t)\ud B^i_t +{\K^{(N)}_t(X^i_t) \circ (\ud Z_t -
	\frac{h(X^i_t) + \hat{h}^{(N)}_t}{2}\ud t)},\quad X^i_0\overset{\text{i.i.d.}}{\sim} p_0,
\end{equation*}  
for $i=1,\ldots N$, where $\{B^i_t:t\geq 0 , 1\leq i\leq N\}$ are mutually independent standard w.p., 
$\hat{h}^{(N)}_t\!:=\!\frac{1}{N}\sum_{i=1}^N h(X^i_t)$ is the empirical approximation of $\hat{h}_t$, and $\K^{(N)}_t$
is the output of an algorithm that approximates the solution to the
Poisson equation~\eqref{eq:Poisson-intro} at each fixed time $t$:
\begin{equation*}
\text{Gain function approximation:}\quad\K^{(N)}_t := \text{Algorithm}(\{X^i_t:1\leq i\leq N\};h).
\end{equation*}
The notation is suggestive of the fact that algorithm is adapted
to the ensemble $\{X^i_t:~1\leq i\leq N\}$ and the function $h$; the density
$p_t(x)$ is not known in an explicit manner.  

Two examples are presented in the sidebars to illustrate the FPF algorithm
in practice: In the first example ``Benefits of feedback'', analytical
and numerical comparisons are provided to show how feedback
control can help ameliorate the CoD.  In the second example ``FPF for SIR
models'' the FPF algorithm is applied to an epidemiological SIR model.

At this point, it is
instructive to specialize FPF to the linear Gaussian case where the
solution of the Poisson equation is explicitly known.

\subsection{FPF for Linear Gaussian setting}
Suppose $a(x)=Ax$, $h(x)=H x$, and $p_t$
is a Gaussian density with mean $\bar{m}_t$ and variance $\bar{\Sigma}_t$.  Then the solution of
the Poisson equation is known in an explicit form~\cite[Sec. D]{yang2016}.  The resulting gain
function is constant and equal to the Kalman gain:
\begin{equation}\label{eq:Kalman-gain}
\K_t(x) \equiv \bar{\Sigma}_tH^\top ,\quad \forall \; x\in \Re^d.
\end{equation}
Therefore, the mean-field process~\eqref{eq:FPF-mean-field} for the
linear Gaussian problem is given as
\begin{equation*}
\ud \bar{X}_t = A \bar{X}_t \ud t + \ud \bar{B}_t + \bar{\Sigma}_tH^\top  (\ud Z_t - \frac{H\bar{X}_t+H\bar{m}_t}{2}\ud t),\quad \bar{X}_0 \sim p_0.
\end{equation*}

Given the explicit form of the gain function~\eqref{eq:Kalman-gain}, the empirical approximation of the gain is simply $\K_t^{(N)}=\Sigma_t^{(N)}H^\top$ where $\Sigma_t^{(N)}$ is the empirical covariance of the particles. Therefore, the evolution of the particles is:
\begin{equation}
\ud X^i_t = AX^i_t \ud t + \ud B^i_t +{\K^{(N)}_t  (\ud Z_t -
	\frac{HX^i_t + H m^{(N)}_t}{2}\ud t)},\quad X^i_0\overset{\text{i.i.d.}}{\sim} p_0,\quad\label{eq:FPF-linear}
\end{equation}
for $i=1.\ldots,N$, where $m_t^{(N)}$ is the empirical mean of the particles. The empirical quantities are computed as:
\begin{align*}
m^{(N)}_t&:=\frac{1}{N}\sum_{i=1}^N X^i_t,\quad\Sigma^{(N)}_t
:=\frac{1}{N-1}\sum_{i=1}^N (X^i_t-m^{(N)}_t)(X^i_t-m^{(N)}_t)^\top.
\end{align*}
The linear Gaussian FPF~\eqref{eq:FPF-linear} is identical to the
square-root form of the ensemble
Kalman filter (EnKF)~\cite[Eq. 3.3]{Reich-ensemble}.

The main difficulty in implementing an FPF in general settings is the
gain function approximation.  Two algorithms for this problem are
presented in the following section.

\section{Gain function approximation}\label{sec:nonlinear}

\begin{figure}
	\centering
	\includegraphics[width=0.6\hsize]{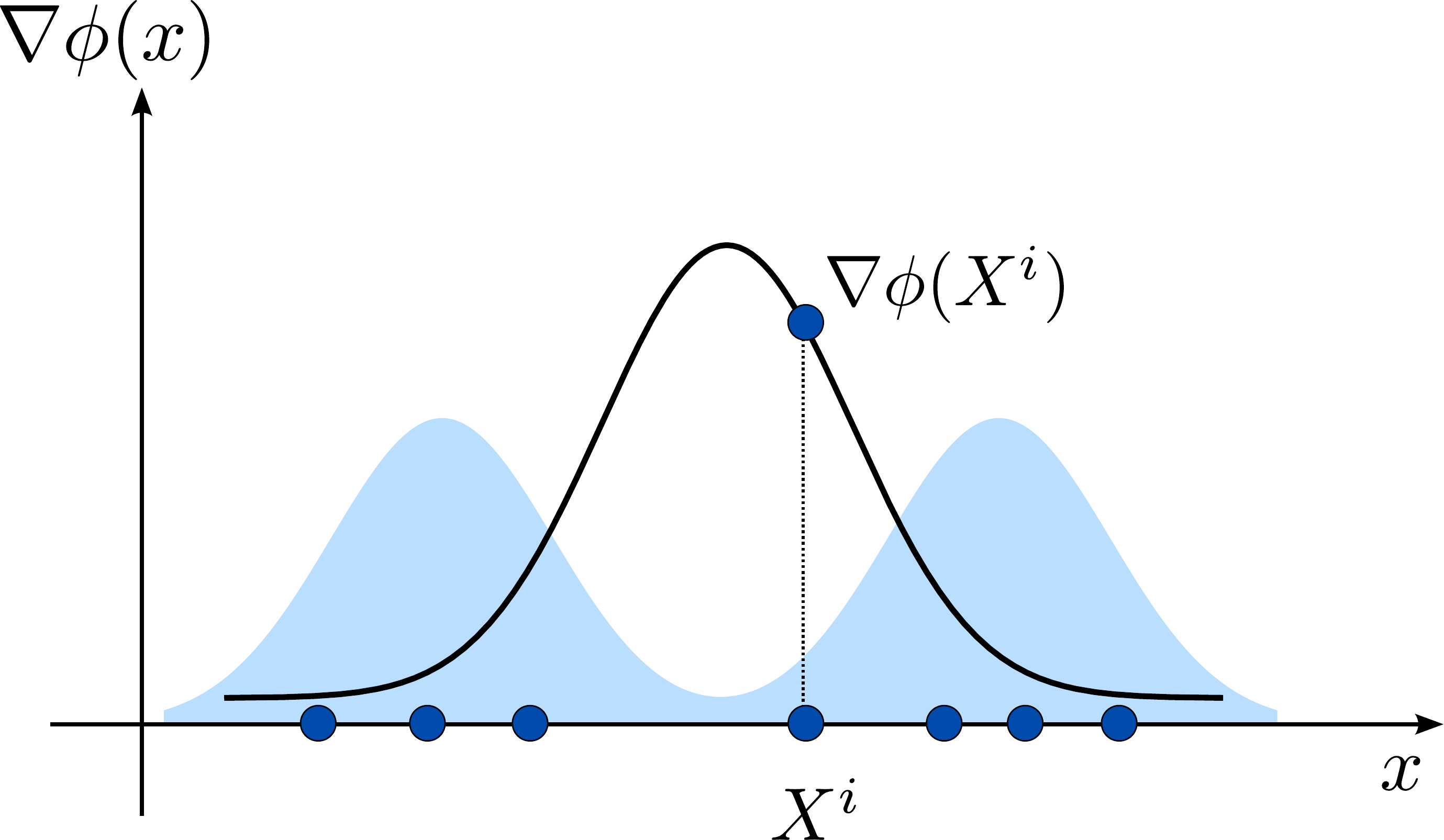}
	\caption{Gain function approximation problem in the feedback particle filter. The exact gain function
          $\K(x)= \nabla \phi(x)$ where $\phi$ solves the Poisson
          equation~\eqref{eq:Poisson-intro}. The numerical problem is
          to approximate $\left. \nabla
            \phi(x)\right|_{x=X^i}$ using only the particles $\{X^i:
          1\leq i\leq N\}$ sampled from density $\rho$ (depicted as
          shaded region).}
	\label{fig:gain-function-approx}
\end{figure}

The exact gain function is a solution of the Poisson equation~\eqref{eq:Poisson-intro}.  
In practice, this problem is solved numerically:
\begin{align*}
    \text{Input:} \quad& \text{samples $\{X^i:1\leq i\leq N\} \overset{\text{i.i.d.}}{\sim} \rho $, observation function $h(\cdot)$},\\
    \text{output:}\quad& \text{gain function $\{\K(X^i): 1\leq i \leq N\}$},
\end{align*}
where in filtering $\rho$ is the (posterior) density at time $t$.  The
explicit dependence on time $t$ is suppressed in this section. The
problem is illustrated in Fig.~\ref{fig:gain-function-approx}. 

\subsection{Constant gain approximation}
\begin{figure}
	\centering
	\includegraphics[width=0.6\hsize]{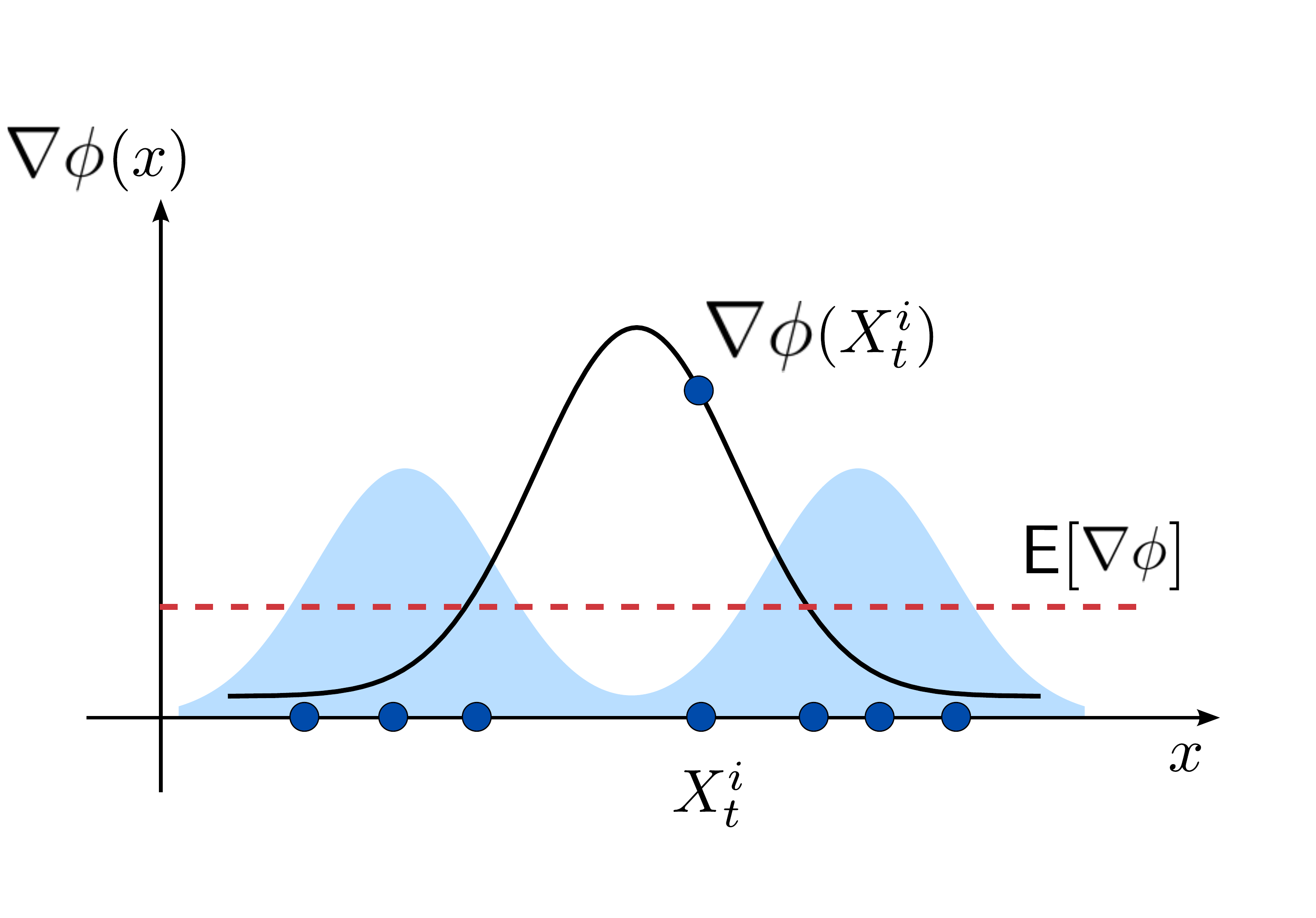}
	\caption{Constant gain approximation in the feedback particle filter. The gain function is approximated by its expected value according to~\eqref{eq:const-gain-approx}.}
	\label{fig:const-gain}
\end{figure}

The simplest approximation is the {\em constant gain
	approximation} formula where the gain $\K_t$ is approximated by
its expected value (which represents the best 
least-square approximation of the gain by a constant).  Remarkably, the
expected value admits a closed-form expression 
which is then readily approximated empirically using the particles:
\begin{equation}
\begin{aligned}
\text{Const. gain approx:}\quad 
\Expect [\K_t (X_t)|\clZ_t] &= \int_{\Re^d} (h(x)-\hat{h}_t)\;
x \; p_t(x) \ud x \\&\approx \frac{1}{N}\sum_{i=1}^N\; (h(X^i_t)-\hat{h}^{(N)}_t) \; X^i_t.
\end{aligned}\label{eq:const-gain-approx}
\end{equation}
(See Figure~\ref{fig:const-gain} for an illustration of the constant
gain approximation.) 
With the constant gain approximation, the FPF algorithm simplifies to
an EnKF algorithm~\cite{TaghvaeiASME2017}. The constant gain
formula~\eqref{eq:const-gain-approx} was known in the EnKF literature
prior to the FPF derivation~\cite{evensen2006,Reich-ensemble}.

There have been a number of studies to improve upon this
formula~\cite{yang2016,yang2013feedback,berntorp2016,matsuura2016suboptimal,Sean_CDC2016,radhakrishnan2018feedback,berntorp2018comparison}.
In the following, we describe the diffusion map approximation which
appears to be the most promising approach in general settings.

\subsection{Diffusion map-based algorithm}

The notation $e^{\epsilon \Delta_\rho}$ is used to denote the
semigroup associated with the probability weighted Laplacian 
$\Delta_\rho$~\cite{bakry2013}.  As explained in the accompanying
sidebar ``Poisson equation and its approximations'' (and more fully in~\cite{taghvaei2019diffusion}), the Poisson
equation~\eqref{eq:Poisson-intro} is equivalently expressed as the fixed-point equation:
\begin{align}
\phi = e^{\Delta_{\rho} \epsilon} \phi+
\int_0^\epsilon e^{\Delta_{\rho} s} (h-\hat{h}) \ud s. \label{eq:Poisson-semigroup} 
\end{align}
where $\epsilon>0$ is arbitrary.  For small values of $\epsilon$,
there is a well known approximation of the exact semigroup
$e^{\Delta_{\rho} \epsilon}$ in terms of the so-called diffusion map:  
\begin{equation*}
\Teps f(x) := \frac{1}{n_\epsilon(x)}\int_{\Re^d}\frac{g_\epsilon(x-y)}{\sqrt{\int g_\epsilon(y-z)\rho(z)\ud z}}f(y)\rho(y)\ud y,
\end{equation*} 
where $g_\epsilon(x):=e^{-\frac{|x|^2}{4\epsilon}}$ is the Gaussian
kernel in $\Re$ and $n_\epsilon(x)$ is the normalization factor chosen
so that $\int \Teps 1(x) \ud x = 1$~\cite{coifman}. 

It is straightforward to approximate the diffusion map empirically in
terms of the particles:
\begin{equation*}
\TepsN f(x) =  \frac{1}{n_\epsilon^{(N)}(x)}\sum_{i=1}^N \frac{g_\epsilon(x-X^i)}{\sqrt{\sum_{j=1}^N g_\epsilon(X^i-X^j)}}f(X^i),
\end{equation*} 
where $n_\epsilon^{(N)}(x)$ is the normalization factor.

Upon approximating the fixed-point
equation~\eqref{eq:Poisson-semigroup}  using the empirical
approximation $\TepsN$ for $e^{\Delta_{\rho} \epsilon}$, one obtains the diffusion
map-based algorithm.  The algorithm is summarized in the sidebar
``Diffusion map-based algorithm for gain function approximation".

 \subsection{Error: bias variance trade-off}
\begin{figure}[t]
	\centering
	\begin{tabular}{cc}
		\begin{subfigure}{.5\textwidth}
			\includegraphics[width=1.0\columnwidth]{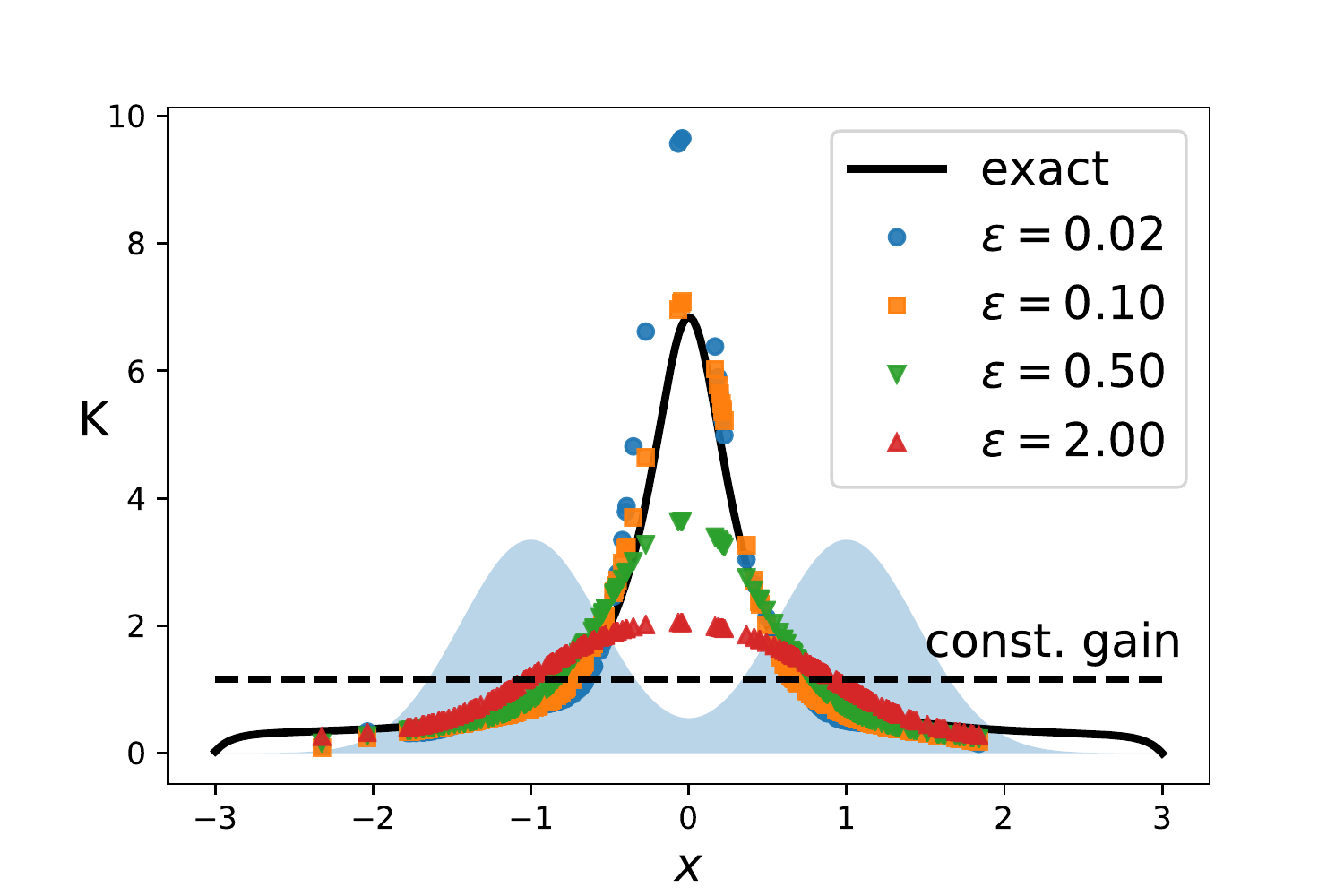}
			\caption{Approximate gain function compared to
				the exact gain function (solid line),}
		\end{subfigure}
		& 
		\begin{subfigure}{.5\textwidth}
			\includegraphics[width=1.0\columnwidth]{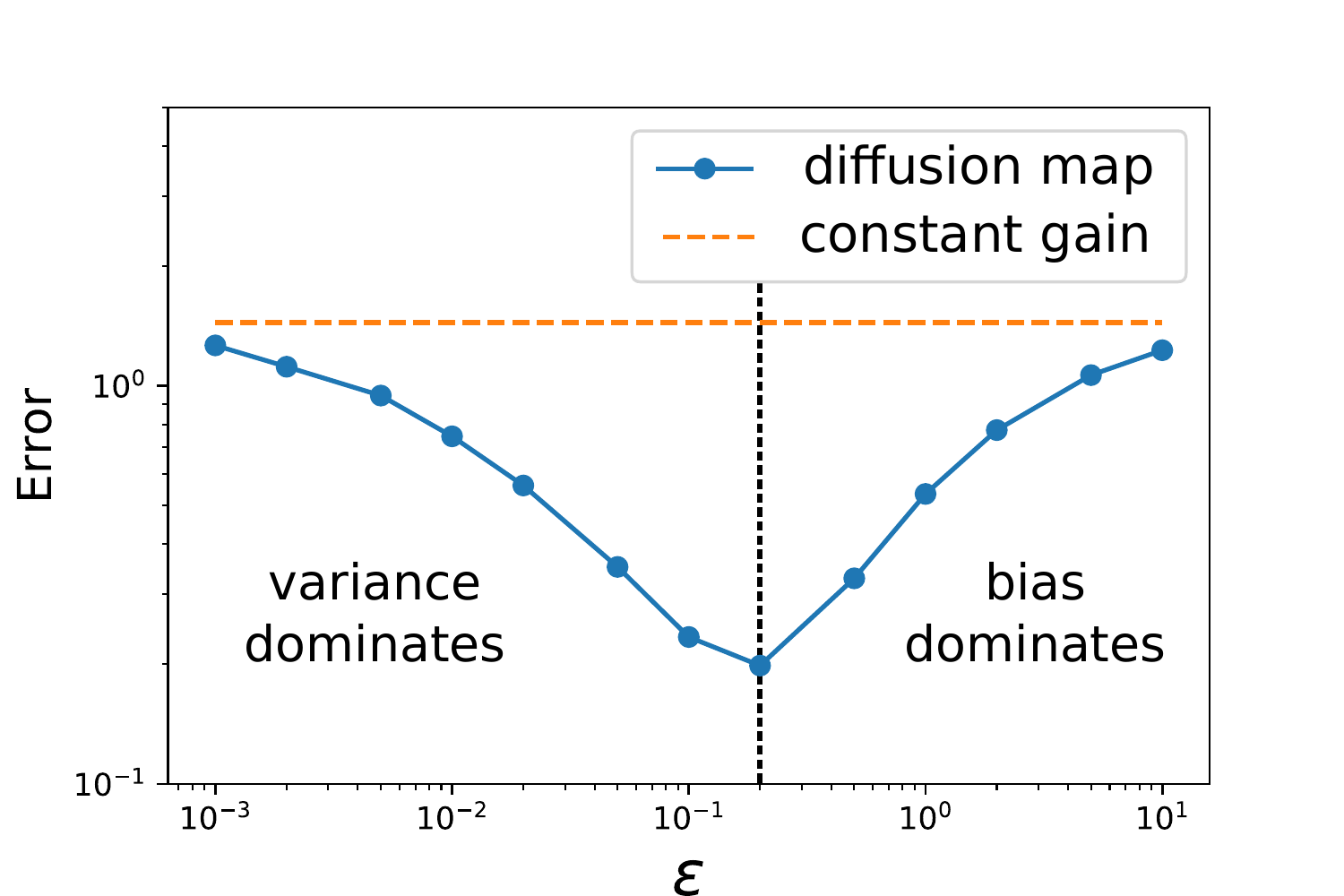}
			\caption{The root mean squared error as a function of the parameter
				$\epsilon$.}
		\end{subfigure}
	\end{tabular}
	\caption{Bias variance trade-off
          in diffusion map-based gain function approximation. (a)
 The dashed line is the constant
		gain solution~\eqref{eq:const-gain-approx}. As
                $\epsilon \to \infty$, the diffusion map gain converges
                to the constant gain.  (The shaded area in the background 
		is the  density function $\rho$ taken as sum of two Gaussians
		 ${\cal N}(-1,\sigma^2)$ and ${\cal N}(+1,\sigma^2)$
                 with $\sigma^2=0.2$.  The exact gain function $\K(x)$
                 is computed for $h(x)=x$ by using an integral
                 formula~\cite[Eq. 4.6]{taghvaei2019diffusion}.) (b)
                 The r.m.s.e. is computed as an empirical
                 approximation of~\eqref{eq:MC_error} by averaging
                 over $1000$ simulations for $N=200$ particles.}
	\label{fig:kernel-approx}
\end{figure}

 The error in diffusion map approximation comes from two sources: (i)
 the bias error due to the diffusion map approximation of the
 semigroup; (ii) the variance error due to empirical approximation in
 terms of particles. The error is analyzed
 in~\cite{taghvaei2019diffusion} where it is shown that
 \begin{equation}
\text{r.m.s.e} = \left(\Expect[ \frac{1}{N}\sum_{i=1}^N | {\sf
  K}(X^i) - {\sf K}_{\text{exact}}(X^i) |^2] \right)^\half\leq \underbrace{O(\epsilon)}_{\text{bias}} + \underbrace{O(\frac{1}{\epsilon^{1+\frac{d}{2}}N^{\frac{1}{2}}})}_{\text{variance}}.
\label{eq:MC_error}
\end{equation}
The error due to bias converges to zero as $\epsilon \to 0$ and the
error due to variance converges to zero as $N \to \infty$.  There is
trade-off between the two errors: To reduce bias, one must reduce
$\epsilon$.  However, for any fixed value of $N$, one can reduce
$\epsilon$ only
up to a point where the variance starts increasing.  
The bais-variance trade-off is illustrated
Fig.~\ref{fig:kernel-approx}:  If $\epsilon$ is large, the error
due to bias dominates, while if $\epsilon$ is small, the error due to
variance dominates.

As a final point, there is a remarkable and somewhat unexpected 
relationship between the diffusion map and the constant gain
approximations.  In particular, in the limit as
$\epsilon \to \infty$, the diffusion map gain converges to the
constant gain.  This suggests a systematic procedure to improve 
upon the constant gain by de-tuning the value of
$\epsilon$ away from the [$\epsilon
= \infty$] limit.  For any fixed $N$, a finite value of $\epsilon$ is chosen to
minimize the r.m.s.e. according to the bias variance trade-off.  Based
on this, a rule of thumb for choosing
the $\epsilon$ value appears in~\cite[Remark 5.1]{taghvaei2019diffusion}.

\section{Some Final Remarks}

In the past decade, the coupling perspective to data assimilation
problems has been enormously valuable with outstanding theoretical
contributions and application impact.  Given the
limited scope of the present article with its narrow focus on the FPF
algorithm, it is not possible to do justice to the depth and breadth of this
exciting new area in one article.  The reader is referred to the
recent monograph~\cite{reich2015probabilistic} and the tutorial style
review article~\cite{reich2019data} for an excellent 
introduction to the subject. 
 
A few important remarks are also necessary at this point:  The 
continuous-time formulation is stressed in this paper for the reasons
of mathematical elegance and beauty.  In practice, discrete-time 
formulations are much more common.  The coupling
viewpoint also applies to these settings~\cite{reich2015probabilistic}
and was in fact also used
in the paper to introduce the main ideas.  Next, optimal couplings are
almost always difficult to compute.  Most popular forms of couplings
used in practice are sub-optimal.  This is true for the classical EnKF
algorithm and also of the FPF algorithm.  (A discussion and exactness
and optimality for FPF appears in the sidebar.)  
As a final point, closely related to the coupling viewpoint is the
gradient flow interpretation of the Bayes' update formula --
see~\cite{laugesen15} for an 
FPF-specific exposition and
also~\cite{halder2018gradient,garbuno2020interacting}
for related algorithms.

There are several directions for future work: It is an open problem to
fully carry out the stability and error analysis of the finite-$N$
FPF particle system with the diffusion map-based gain function
approximation.  It will be very useful to be able to characterize the
CoD in these general settings.  It is also of interest to construct
optimization type formulations that directly yield a finite-$N$
algorithm without the need for empirical approximation as an
intermediate step. Such constructions may lead to better error
properties by design. Finally, apart from the optimal
transportation formulation stressed in this paper, one may consider
alternative approaches for control design.  One possible direction is
based on the Schr\"odinger bridge
problem~\cite{chen2016relation,reich2019data}.

\bibliographystyle{IEEEtran}
\bibliography{TAC-OPT-FPF,fpfbib,ref,fpfbib2,Optimization,meanfield,meanfield_v2,acc-flow}

\processdelayedfloats 

\sidebars 

\clearpage
\newpage

\section{Summary}
\label{sidebar:summary}

Feedback particle filter (FPF) is a Monte-Carlo (MC) algorithm 
to approximate the solution of a stochastic filtering 
problem.  In contrast to conventional particle filters, the 
Bayesian update step in FPF is implemented via a mean-field 
type feedback control law.

The objective for this paper is to situate the development 
of FPF and related controlled interacting particle 
system algorithms within the framework of optimal 
transportation theory.  Starting from the simplest setting 
of the Bayes' update formula, a coupling viewpoint is introduced to 
construct particle filters.  It is shown that the conventional 
importance sampling resampling particle filter implements 
an independent coupling.  Design of optimal 
couplings is introduced first for the simple Gaussian 
settings and subsequently extended to derive the FPF 
algorithm.  The final half of the paper provides a 
review of some of the salient aspects of the FPF algorithm 
including the feedback structure, algorithms for gain 
function design, and comparison with conventional particle 
filters.  The comparison serves to illustrate the benefit of feedback in
particle filtering.     

\newpage
\clearpage 
\section{Summary of the linear Gaussian  example}
The formulae for the linear Gaussian example are summarized as follows:
\begin{align*}
\text{Prior:}&\quad \text{Gaussian } \NN(m_0,\Sigma_0)\\
\text{Observation model:}&\quad Y = HX + W\\\
\text{Likelihood function:}&\quad l(x) = \exp(-\frac{|y-Hx|^2}{2})\\
\text{Posterior:}&\quad \text{Gaussian } \NN(m_1,\Sigma_1)\\
\text{Optimal transport map:}&\quad T(x) = F(x-m_0) + m_{1} 
\\
\text{Mean-field process:}&\quad \Xbar_1 = F(\Xbar_0-m_0) + m_0 + \K(Y-Hm_0)
\\
\text{Particle system:}&\quad X^i_{1} = F^{(N)}(X^i_0-\mN_0) + \mN_0 + \KN(Y-H\mN_0),\quad\text{for}\quad i=1,\ldots,N
\end{align*}

\newpage
\clearpage
\section{Optimal transport construction of stochastic processes}
\subsection{Deterministic path}
Let ${\mathcal P}_2(\Re^d)$ be the space of everywhere positive probability densities
on $\Re^d$ with finite second moment.  Given a smooth path
$\{p_t \in \mathcal P_2(\Re^d):~t\geq 0\}$ the problem is to
construct a stochastic process $\{\Xbar_t;~t\geq 0\}$ such that the
probability density of $\Xbar_t$, denoted as $\bar{p}_t$, equals $p_t$ for all
$t\geq 0$. The exactness condition is expressed as
\begin{equation}\label{eq:exactness}
\bar{p}_t = p_t,\quad \forall \;\; t\geq 0.
\end{equation}
Now, there are infinitely many stochastic processes that satisfy
the exactness condition. This is because the exactness condition
specifies only the one-time marginal distribution which is clearly not enough
to uniquely identify the stochastic process, e.g., the two-time joint
distributions are not specified.  A unique choice is made by
prescribing an additional optimality criterion based on the optimal
transportation theory.  

To make these considerations concrete, assume that the given path
$\{p_t:~t\geq 0\}$ evolves according to the PDE
\begin{equation*}
\frac{\partial p_t}{\partial t} = \mathcal V(p_t),
\end{equation*}
where $\mathcal V(\cdot)$ is an operator (e.g., the Laplacian) that acts
on probability densities.  (This necessarily restricts the
operator $\mathcal V$, e.g., $\int \mathcal V(p)(x) \ud x = 0$ for all
$p\in \mathcal P_2(\Re^d)$.) 
The following model is assumed for the process $\{\Xbar_t:t\geq 0\}$:
\begin{equation}\label{eq:Xbar-u}
\frac{\ud}{\ud t} \Xbar_t  = u_t(\Xbar_t),\quad \Xbar_0 \sim p_0,
\end{equation}
where $u_t(\cdot)$ is a control law that needs to be designed. Using
the continuity equation, the exactness condition~\eqref{eq:exactness}
will be satisfied if
\begin{equation}\label{eq:u-const}
-\nabla \cdot(\bar{p}_t u_t) = \mathcal V(\bar{p}_t),\quad \forall\;\; t \geq 0.
\end{equation}




The non-uniqueness issue is now readily seen: The first-order
PDE~\eqref{eq:u-const} admits infinitely many solutions. 
A unique solution $u_t$ is picked by optimizing the coupling between $\Xbar_t$ and
$\Xbar_{t+\Delta t}$ in the limit as $\Delta t \to 0$.  The leading
term in the transportation cost $\Expect[|X_{t+\Delta t}- X_|^2]$ is
of order $O(\Delta t^2)$ whereby
\begin{equation*}
\lim_{\Delta t \to 0}\frac{1}{\Delta t^2}\Expect[|X_{t+\Delta t} - X_t|^2]=\int_{\Re^d} |u_t(x)|^2\bar{p}_t(x) \ud x.
\end{equation*}
Therefore, for each fixed $t\in[0,1]$, the control law $u_t$ is
obtained 
by solving the constrained optimization problem 
\begin{equation}\label{eq:opt-u}
\min_{u_t} \int_{\Re^d} |u_t(x)|^2\bar{p}_t(x) \ud x,\quad \text{s.t}\quad    -\nabla \cdot(\bar{p}_t u_t) = \mathcal V(\bar{p}_t).
\end{equation}
The cost is the infinitesimal form of the $L^2$-Wasserstein
distance and the constraint expresses the exactness condition.  

By a standard calculus of variation argument, the solution of the
optimization problem~\eqref{eq:opt-u} is obtained as $u^*_t=\nabla
\phi_t$ where $\phi_t$ solves the second-order PDE $-\nabla
\cdot(\bar{p}_t \nabla \phi_t) = \mathcal V(\bar{p}_t)$. 
The resulting stochastic process $\Xbar_t$ evolves according to
\begin{align*}
&\frac{\ud \Xbar_t}{\ud t} = \nabla \phi_t(\Xbar_t),\quad \Xbar_0\sim p_0,\\
&\phi_t \text{ solves the PDE } -\nabla \cdot(\bar{p}_t \nabla \phi_t) = \mathcal V(\bar{p}_t).
\end{align*}


As a concrete example, suppose the given path is a solution of the heat equation
$\frac{\partial p_t}{\partial t} = \Delta p_t$. (So $\calV(\cdot)$ is the
Laplacian operator.)   The solution of the second-order PDE is easily
obtained as $\phi_t = \log(\bar{p}_t)$.  The optimal transport process
$\Xbar_t$ then evolves according to
\begin{equation*}
\frac{\ud}{\ud t} \Xbar_t = -\nabla \log(p_t(\Xbar_t)),\quad \Xbar_0\sim p_0.
\end{equation*}
This process should be compared to the SDE
\begin{equation}\label{eq:Xbar-heat-s}
\ud X_t = \ud B_t,\quad X_0\sim p_0,
\end{equation}
where $\{B_t:t\geq 0\}$ is a w.p..  The SDE~\eqref{eq:Xbar-heat-s} is a
well-known stochastic coupling whose one-point marginal evolves
according to the solution of the heat equation.


\subsection{Stochastic path}
In the filtering problem, the path of the posterior probability
densities is stochastic (because it depends upon random
observations $\{Z_t:t\geq 0\}$). Therefore, the preceding discussion
is not directly 
applicable.  Suppose the stochastic path
$\{p_t(\cdot) \in \mathcal P_2(\Re^d):~t\geq 0\}$ is governed by a
stochastic PDE
\begin{equation*}
\ud p_t = \mathcal H(p_t) \ud I_t,
\end{equation*}
where $\mathcal H(\cdot)$ is an operator that acts on probability
densities 
and $\{I_t:t\geq 0\}$ is a w.p.. 

Consider the following SDE model:
\begin{equation*}
\ud \Xbar_t = u_t(\Xbar_t)\ud t + \K_t(\Xbar_t) \ud I_t,\quad \Xbar_0 \sim p_0,
\end{equation*}
where, compared to the deterministic form of~\eqref{eq:Xbar-u}, 
an additional stochastic term is now included. The problem is to
identify control laws $u_t(\cdot)$ and $\K_t(\cdot)$ such that
the conditional distribution of $\Xbar_t$ equals $p_t$. This exactness
condition, counterpart of~\eqref{eq:u-const}, now is
\begin{subequations}
	\begin{align}
	&-\nabla \cdot(\bar{p}_t\K_t) = \mathcal H(\bar{p}_t),\\
	&-\nabla \cdot(\bar{p}_t u_t) + 
	\frac{1}{2}(\nabla \cdot(\bar{p}_t\K_t) \K_t + \bar{p}_t\K_t\nabla \K_t) = 0.\label{eq:const-u}
	\end{align}
\end{subequations}
These equations are obtained by writing the time-evolution of the
conditional probability density of
$\Xbar_t$~\cite[Prop. 1]{yang2016}.  As in the deterministic setting,
the solution is not unique.  

The unique optimal control law is obtained by requiring that the
coupling between $\Xbar_t$ and $\Xbar_{t+\Delta t}$ is optimal in the
limit as $\Delta t \to 0$.  
In contrast to the deterministic setting, the leading term in the
transportation cost $\Expect[|X_{t+\Delta t}- X_|^2]$ is of order
$O(\Delta t)$ whereby
\begin{equation}\label{eq:opt-problem-K-1}
\lim_{\Delta t \to 0}\frac{1}{\Delta t}\Expect[|X_{t+\Delta t} - X_t|^2]=\int_{\Re^d} |\K_t(x)|^2\bar{p}_t(x) \ud x.
\end{equation}
Therefore, for each fixed $t\in[0,1]$, the control law $\K_t$ is
obtained 
by solving the constrained optimization problem 
\begin{equation}\label{eq:opt-problem-K-2}
\min_{\K_t} \int_{\Re^d} |\K_t(x)|^2\bar{p}_t(x) \ud x,\quad \text{s.t}\quad    -\nabla \cdot(\bar{p}_t \K_t) = \mathcal H_t(\bar{p}_t).
\end{equation}
As before, the solution of the optimization
problem~\eqref{eq:opt-problem-K-2} is given by 
$\K^*_t=\nabla \phi_t$ where $\phi_t$ solves the second-order PDE
$-\nabla \cdot(\bar{p}_t \nabla \phi_t) = \mathcal H(\bar{p}_t)$. 

It remains to identify the control law for $u_t$.  For this purpose,
the second-order term in the infinitesimal Wasserstein cost is used: 
\begin{align*}
\lim_{\Delta t \to 0}\frac{1}{\Delta t^2}\left(\Expect[|X_{t+\Delta t} - X_t|^2] - \Delta t \int_{\Re^d} |\K^*_t|^2\bar{p}_t\ud x \right) = \int_{\Re^d} |u_t|^2\bar{p}_t \ud x.
\end{align*}
The righthand-side is minimized subject to the
constraint~\eqref{eq:const-u}.  Remarkably, the optimal solution is
expressed as 
\[
u^*_t = -\frac{1}{2\bar{p}_t}\mathcal{H}(\bar{p}_t)\nabla \phi_t +
\frac{1}{2}\nabla^2 \phi_t \nabla \phi_t + \xi_t,
\] 
where $\xi_t$ is the (unique such) divergence free vector field
(i.e., $\nabla \cdot(p_t\xi_t)=0$) such that $u_t$ is of a gradient form.
The resulting optimal transport process is
\begin{align}\label{eq:OT-process-stoch}
&\ud \Xbar_t = \nabla \phi_t(\Xbar_t) \circ (\ud I_t - \frac{1}{2\bar{p}_t}\mathcal{H}(\bar{p}_t)\ud t) + \xi_t(\Xbar_t)\ud t ,\quad \Xbar_0 \sim p_0.
\end{align}
It is also readily shown that the process $\{\bar{X}_t:t\geq 0\}$ is
in fact exact for any choice of divergence free vector field
$\xi_t$.  The most convenient such choice is obtained by
simply setting $\xi_t\equiv 0$.  The resulting filter is exact and
furthermore also (infinitesimally) optimal to the first-order
(see~\eqref{eq:opt-problem-K-1}).

For the special case of the nonlinear filtering problem,
$\mathcal{H}(p) = (h-\hat{h})p$ where $\hat{h} = \int h(x) p(x) \ud
x$ and $\ud I_t = \ud Z_t - \hat{h}_t\ud t$ is the increment of the
innovation process.  The optimal transport stochastic process
\eqref{eq:OT-process-stoch} is then given by the Stratonovich form 
\begin{align*}
&\ud \Xbar_t = \nabla \phi_t(\Xbar_t) \circ (\ud Z_t - \frac{1}{2}(h+\hat{h}_t)\ud t) + \xi_t(\Xbar_t)\ud t ,\quad \Xbar_0 \sim p_0.
\end{align*}
The control law in the FPF algorithm~\eqref{eq:FPF-mean-field}
represents the particular sub-optimal choice $\xi_t\equiv 0$.


\newpage
\clearpage
 \section{Benefit of Feedback}
In this sidebar, we consider a simple example to highlight the phenomena of curse of dimensionality (CoD) in particle filters and provide comparison with FPF to see how the curse is mitigated by using feedback control.  The example we consider is as follows: 
\begin{subequations}\label{eq:filter-problem-example}
	\begin{align}
	\ud X_t &= 0, \quad X_0 \sim \NN(0,\sigma_0^2I_d),\\ 
	\ud Z_t &= X_t \ud t + \sigma_w \ud W_t,
	\end{align}
\end{subequations}
for $t\in[0,1]$, where $X_t$ is a $d$-dimensional process, $\sigma_w,\sigma_0>0$ and $I_d$ is a $d\times d$ identity matrix. The posterior distribution at time $t=1$ is a Gaussian distribution $\NN(m_1,\Sigma_1)$ with mean $m_1=\frac{\sigma_0^2}{\sigma_0^2+ \sigma_w^2}$ and variance $\Sigma_1 = \frac{\sigma_0^2\sigma_w^2}{\sigma_0^2+ \sigma_w^2}I_d$. 

We consider the following MC approaches to approximate the posterior distribution:
\begin{enumerate}
	\item SIR particle filter (PF): Sample $\{X^i_0:1\leq i\leq N\}$ from the initial distribution.  Form the weighted distribution and generate new samples from the weighted distribution. 
	\begin{equation}\label{eq:PF-Est} 
	X^i_1 \sim \sum_{i=1}^N w_i\delta_{X^i_0},\quad w_i = \frac{e^{-\frac{|Z_1-X^i_0|^2}{2\sigma_w^2}}}{\Expect[e^{-\frac{|Z_1-X^i_0|^2}{2\sigma_w^2}}|\clZ_1]},\quad X^i_0 \sim \NN(0,\sigma_0^2I_d).
	\end{equation}
	\item Feedback particle filter (FPF):  Simulate the particles according to 
	\begin{equation}\label{eq:FPF-Est}
	\ud X^i_t = \frac{1}{\sigma_w^2}\SigN_t (\ud Z_t - \frac{X^i_t + \mN_t}{2} \ud t)  ,\quad X^i_0 \sim \NN(0,\sigma_0^2I_d). 
	\end{equation}
	for $t\in[0,1]$, 
	where $\mN_t$ is the empirical mean of the particles, and $\SigN_t$ is the empirical variance of the particles.  
\end{enumerate} 

We compare the mean squared error (m.s.e) in approximating the exact
conditional mean $m_1$ of $X_1$. The m.s.e is defined as
\begin{equation*}
\text{m.s.e} := \Expect[\frac{1}{N}\sum_{i=1}^N |X^i_1 - m_1|^2].
\end{equation*}
\

In~\cite{Taghvaei2019AnOT}, the following is proved:

\begin{proposition}\label{prop:importance-sampling}
	Consider the filtering problem~\eqref{eq:filter-problem-example} with
	state dimension $d$.  
	Then
	\begin{romannum}
		\item For the PF~\eqref{eq:PF-Est} 
		\begin{equation*}
		\text{m.s.e}_{\overline{\text{PF}}}(f) = \frac{\sigma^2}{N}\left(3(2^d) - \frac{1}{2}\right) \geq \frac{\sigma^2}{N}2^{d+1}.
		\end{equation*}
		\item For the FPF~\eqref{eq:FPF-Est}
		\begin{equation*}
		\text{m.s.e}_{{\text{FPF}}}(f) \leq  \frac{\sigma^2}{N} (3d^2+2d).
		\end{equation*}
	\end{romannum}
\end{proposition}

The result above is consistent with the extensive studies on 
importance sampling-based particle
filters~\cite{bickel2008sharp,bengtsson08,snyder2008obstacles,rebeschini2015can}.
In these papers, it is shown that if $\frac{\log N \log d}{d}\to 0$
then the largest importance weight $\max_{1\le i \le N} w^i \to 1$ in
probability.  Consequently, in order to prevent the weight collapse,
the number of particles must grow exponentially with the dimension.  
This phenomenon is referred to as the curse of dimensionality (CoD)
for the particle filters.

\begin{figure}
	\centering
	\begin{tabular}{cc}
		\begin{subfigure}{.5\textwidth}
			\includegraphics[width = 0.95\hsize]{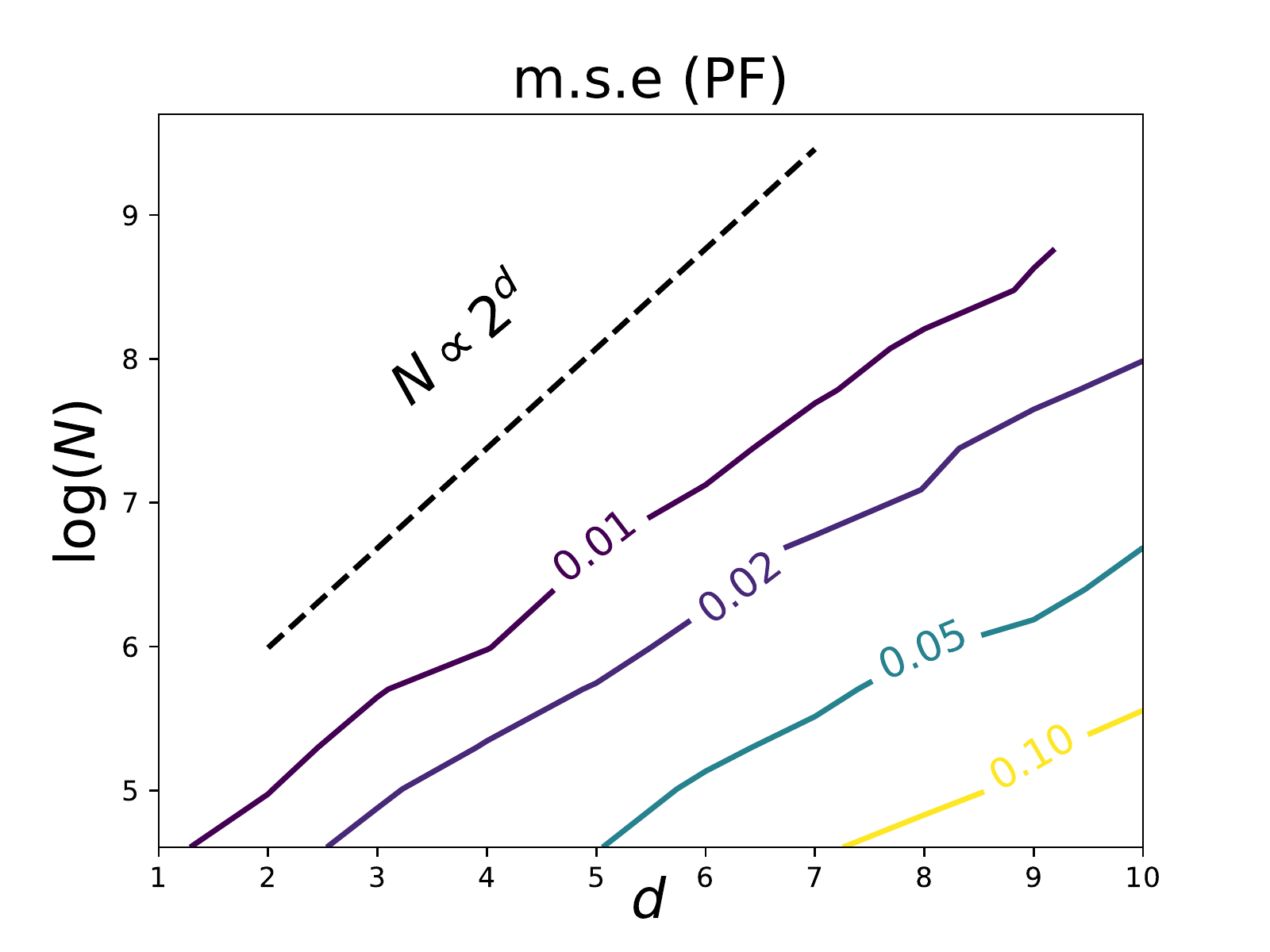}
			\caption{importance
				sampling particle filter (PF)~\eqref{eq:PF-Est}}
			\label{fig:error-PF}
		\end{subfigure}
		&
		\begin{subfigure}{.5\textwidth}
			\includegraphics[width = 0.95\hsize]{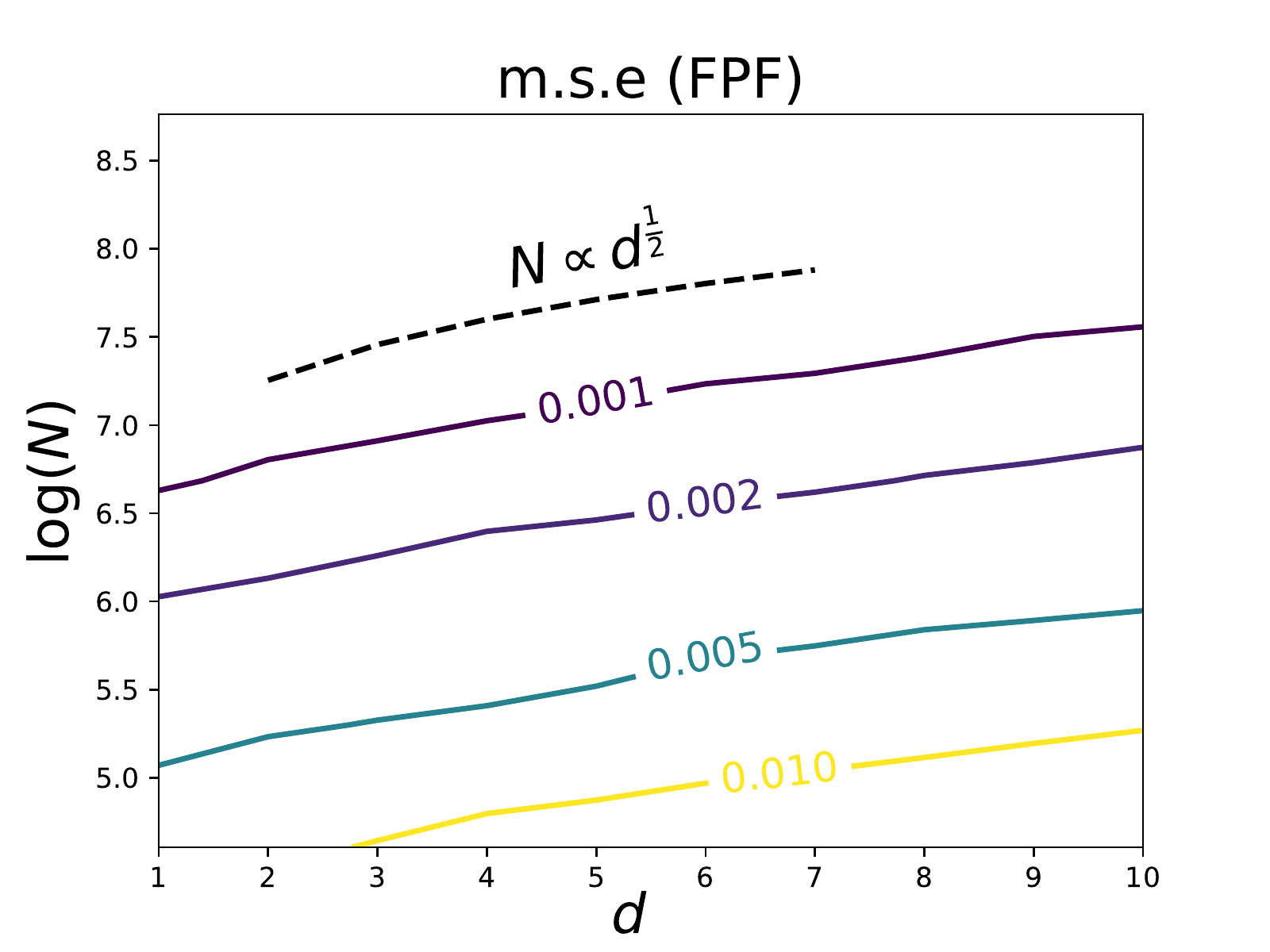}
			\caption{feedback particle filter (FPF)~\eqref{eq:FPF-Est}}
			\label{fig:error-FPF}
		\end{subfigure}
	\end{tabular}
	\caption{Overcoming the curse of dimensionality of particle filters. The solid lines correspond to the level sets of the mean squared error for the filtering
		problem~\eqref{eq:filter-problem-example}.
		In order to have the same error, the PF  requires the number of samples $N$ to grow exponentially with the dimension $d$, whereas the growth using the FPF for this numerical example is $O(d^\half)$.  
	} 
	\label{fig:error-PF-FPF}
\end{figure}

A numerical comparison of the m.s.e. as a function of $N$ and $d$ is
depicted in Figure~\ref{fig:error-PF-FPF}-(a)-(b). The expectation is
approximated by averaging over $M=1000$ independent simulations.  It
is observed that, in order to have the same error, the importance
sampling-based approach requires the number of samples $N$ to grow
exponentially with the dimension $d$, whereas the growth using the FPF
for this numerical example is $O(d^\half)$.  
The scaling with dimension depicted in
Figure~\ref{fig:error-PF-FPF}~(b) suggests that the $O(d^2)$ bound for
the m.s.e in the linear FPF is loose.  This is because of
the conservative nature of approximations used in deriving the inequality~\cite{Taghvaei2019AnOT}.
The overall conclusions of the study are consistent with other
numerical results reported in the
literature~\cite{surace_SIAM_Review}.

 \newpage
 \clearpage

\processdelayedfloats 

\newpage
\clearpage
\section{Poisson equation and its approximations}

The Poisson equation~\eqref{eq:Poisson-intro} of nonlinear filtering
is a linear PDE.  Its finite-dimensional counterpart is a familiar
linear problem
\begin{equation}
    Ax = b,
\label{eq:PE_finite_dim}
\end{equation}
where $A$ is a $n\times n$ (strictly) positive-definite symmetric
matrix and the righthand-side $b$ is a given $n\times 1$ vector.  The
problem is to obtain the unknown $n\times 1$ vector $x$.  For this purpose, the
following equivalent formulations of the finite-dimensional problem are first
introduced: 
\begin{enumerate}
    \item $x$ is the solution of the weak form
    \begin{equation*}
        y^\top A x = y^\top b,\quad \forall \; y \in \Re^n.
    \end{equation*}
    \item  For  any $t>0$, $x$ is the solution to the fixed-point equation
    \begin{equation*}
        x = e^{-tA} x + \int_0^t e^{-sA} b \ud s.
    \end{equation*}
   \item $x$ is the solution of the optimization problem
   \begin{equation*}
       \min_{x\in \Re^n}~\frac{1}{2}x^\top A x - x^\top b.
   \end{equation*}
\end{enumerate}
When $n$ is large, these formulations are useful to numerically approximate the
solution of~\eqref{eq:PE_finite_dim}:
\begin{enumerate}
\item For each fixed $y\in\Re^n$, the weak form is a single equation.
  By restricting $y$ to a suitable low-dimensional subspace 
  $S\subset\Re^n$, the numer of linear equations is reduced for the
  purposes of obtaining an approximate solution (possibly also in $S$).  
\item The fixed-point equation form is useful because $e^{-tA}$ is a
  contraction for positive-definite $A$.  So, a good initial guess for
  $x$ can readily be improved by using the Banach iteration.  
\item The optimization form is useful to develop alternate (e.g.,
  search type) algorithms to obtain the solution to~\eqref{eq:PE_finite_dim}. 
\end{enumerate}

We turn our attention next to the Poisson
equation~\eqref{eq:Poisson-intro} expressed succinctly as 
\begin{equation*}
    -\Delta_\rho \phi = h-\hat{h},
\end{equation*}
where $\Delta_\rho=\frac{1}{\rho}\nabla \cdot (\rho \nabla)$ is the probability weighted Laplacian.  
Functional analytic considerations require introduction of the
function spaces: $L^2(\rho)$ is the space of square integrable
functions with respect to $\rho$ with inner product $ \langle
f,g\rangle_{L^2} = \int f(x)g(x)\rho(x)\ud x$; $H^1(\rho)$ is the
Hilbert space of functions in $L^2(\rho)$ whose first derivative, defined
in the weak sense, is the also in $L^2(\rho)$; and
$H^1_0(\rho)=\{\psi\in H^1(\rho) | \int \psi(x) \rho (x) \ud x = 0\}$.

These definitions are important because $H^1_0(\rho)$ is the natural
space for the solution $\phi$ of the Poisson
equation~\eqref{eq:Poisson-intro}.  The operator $-\Delta_\rho$ is
symmetric (self-adjoint) and positive definite because 
\begin{equation*}
   - \langle f,\Delta_\rho g \rangle_{L^2} = 
    \langle \nabla f,\nabla g\rangle_{L^2}
    = - \langle \Delta_\rho f,g \rangle_{L^2},\quad \forall f,g\in H_0^1(\rho) .
\end{equation*}
One requires an additional technical condition -- the so-called
Poincar\'e inequality -- to conclude that the operator is in fact 
strictly positive-definite.  Assuming the Poincar\'e inequality holds,
it is also readily shown that $\Delta_\rho^{-1}$ is well defined, i.e.,
a unique solution $\phi\in H^1_0(\rho)$ exists for a given $h\in
L^2(\rho)$~\cite[Thm. 2]{yang2016}.

For the purposes of numerical approximation, entirely analogous to the
finite-dimensional case, the following equivalent formulations of the Poisson
equation are introduced:
\begin{enumerate}
    \item $\phi$ is a solution of the weak form
\begin{equation}
 \langle \nabla \psi, \nabla \phi \rangle_{L^2} = \langle \psi,h-\hat{h}\rangle_{L^2}  \quad \forall\; \psi
 \in H_0^1(\rho).\label{eq:Poisson-weak}
 \end{equation}
     \item $\phi$ is a solution of the fixed-point equation
\begin{equation*}
\phi = e^{t \Delta_\rho} \phi + \int_0^t e^{s\Delta_\rho} (h-\hat{h})\ud s.
 \end{equation*}
 \item $\phi$ is the solution of the optimization problem
 \begin{equation}
\min_{\phi \in H^1_0(\rho)} \frac{1}{2} \langle \nabla \phi, \nabla
\phi \rangle_{L^2} + \langle \phi,h-\hat{h}\rangle_{L^2}.
\label{eq:Poisson-opt}
 \end{equation}
\end{enumerate}

These formulations have been used to develop numerical algorithms for gain
function approximation:
\begin{enumerate}
    \item Instead of $\psi \in H^1_0(\rho)$ in the weak
      form~\eqref{eq:Poisson-weak}, a relaxation is considered whereby
      $\psi \in S = \text{span}\{\psi_1,\ldots,\psi_M\}$, a
      finite-dimensional subspace of $H_0^1(\rho)$. The resulting
      algorithm is referred to as the Galerkin
      algorithm for gain function approximation~\cite{yang2016}.  The
      constant gain formula~\eqref{eq:const-gain-approx} is obtained by considering $S$ to
      be the subspace spanned by the coordinate functions.    

    \item The semigroup $e^{t\Delta_\rho}$ is approximated with the
      diffusion map operator $\Teps$ as described in the main body of
      the paper.  This approximation yields the diffusion map-based
      algorithm for gain function approximation, tabulated in the sidebar.   

    \item The optimization formulation~\eqref{eq:Poisson-opt} is
      useful to explore nonlinear parametrizations of the gain
      function, e.g., using neural networks.  A preliminary
      investigation of this appears in our paper~\cite{olmez2020deep}.
      Related deep learning-inspired techniques for solving PDEs using neural networks appear in~\cite{weinan2018deep}.

\end{enumerate}

\newcommand{\phivec}{{\sf \Phi}}
\clearpage
\newpage
 \section{Diffusion-map based algorithm for gain function approximation}
 \noindent
{\bf Input:}   $\{X^i: 1\leq i\leq N\}$,  $\{h(X^i): 1\leq i\leq N\}$,  kernel bandwidth $\epsilon$

\noindent
{\bf Output:} $\{\K^i: 1\leq i\leq N\}$ 

\begin{enumerate}
    \item  $g_{ij}:=e^{-\frac{|X^i-X^j|^2}{4\epsilon}}$ for $i,j=1$ to $N$\medskip
    \item  $k_{ij}:=\frac{g_{ij}}{\sqrt{\sum_l g_{il}}\sqrt{\sum_l g_{jl}}}$ for $i,j=1$ to $N$\medskip
    \item $d_i= \sum_{j} k_{ij}$ for $i=1$ to $N$\medskip
    \item $\Ten_{ij}:=\frac{k_{ij}}{d_i}$ for $i,j=1$ to $N$\medskip
    \item  $\pi_i=\frac{d_i}{\sum_{j}d_j}$ for $i=1$ to $N$\medskip
    \item  $\hat{\hvec}= \sum_{i=1}^N \pi_jh(X^i)$ \medskip
    \item  $\phivec=(0,\ldots,0)\in \Re^N$\medskip
    \item Solve the fixed point problem $\phivec=\Ten \phivec + \epsilon(\hvec - \hat{\hvec})$ iteratively \medskip
    \item  $r_i = \phivec_i + \epsilon \hvec_i$ for $i=1$ to $N$\medskip
    \item  $s_{ij} = \frac{1}{2\epsilon}\Ten_{ij}(r_j-\sum_{k=1}^N
     \Ten_{ik}r_k)$ for $i,j=1$ to $N$\medskip
    \item  $\K^i = \sum_j s_{ij}X^j$ for $i=1$ to $N$
\end{enumerate}

\newpage
\clearpage

\section{Example: FPF for SIR models}
The basic mathematical model of epidemiological disease propagation is
the SIR ODE model:
\begin{align*}
\dot{S}_t &= -\beta S_tI_t, \\
\dot{I}_t &= \beta S_tI_t - \alpha  I_t,\\
\dot{R}_t & = \alpha I_t,
\end{align*}
where $S_t$, $I_t$ and $R_t$ are the susceptible, infected and
recovered population fractions, respectively, at time $t$.  The parameters $\beta$ and
$\alpha$ are the transmission rate and the recovery rate parameters,
respectively.      
In an epidemic, one observes the number of newly infected over a
time-increment (daily).  For our study, this is modeled as
\begin{equation}\label{eq:obs_SIR}
\ud Z_t = (\beta I_t S_t) \ud t + \sigma_W \ud W_t,
\end{equation}
where $W=\{W_t: t\geq 0\}$ is the standard w.p. and $\sigma_W$ is the
standard deviation (std. dev.) parameter.  
Given the observations, the filtering objective is to estimate the
population sizes and possibly also the model parameters.  In this
study, the recovery rate parameter $\alpha$ is assumed known while the
transmission rate parameter $\beta$ is estimated.  In a filtering
setup, this requires a model which is assumed to be of the following form:
\begin{align*}
\ud \beta_t &= \sigma_B \ud B_t,
\end{align*} 
where $B=\{B_t: t\geq 0\}$ is a standard w.p. and $\sigma_B$ is the
standard deviation parameter.

The model and the filter are simulated using the Euler discretization
scheme for time integration.  The simulation parameters are as follows: time
discretization step-size $\Delta t= 1$; std. dev. for the observation noise $\sigma_W =
0.1$; std. dev. for the process noise $\sigma_B  = 0.1$; initial
distribution $I(0) \sim \text{unif}[0,0.1]$ and $S(0)=1-I(0)$;
recovery rate $\alpha=0.1$; 
the transmission rate $\beta$ is fixed to be $0.1$ but assumed unknown
to the filtering algorithm. The FPF is simulated using $N=100$
particles.
Two gain function approximation algorithms are implemented:  the constant gain approximation and the diffusion map approximation. For the diffusion map approximation, the heuristic $\epsilon = 10 \text{med}(\{|X^i-X^j|^2;1\leq i,j\leq N\}) (\log(N))^{-1}$ is used, where $\text{med}(\cdot)$ denotes the statistical median.
The simulation parameters and their values are tabulated in
Table~\ref{tab:params}.

\begin{figure}[t]
	\centering
	
	\includegraphics[width=0.9\columnwidth]{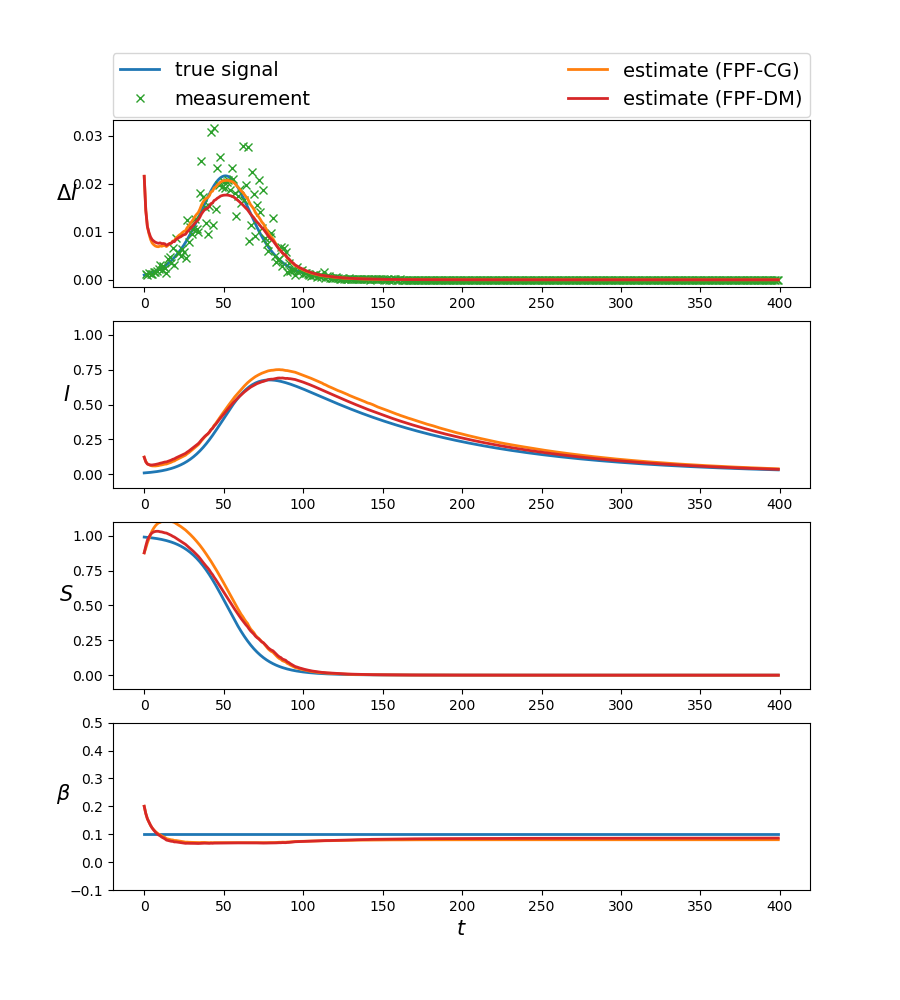}
	
	\caption{Application of the feedback particle filter on the SIR epidemiological model. The observation is the number of new confirmed cases each day $\Delta I$ (depicted in the top figure).  The size of infected population is $I(t)$ and the size of susceptible population is $S(t)$. The infection transmission rate $\beta$ is assumed unknown and it is estimated.  The estimation algorithm is feedback particle filter with constant gain approximation (FPF-CG) and feedback particle filter with the diffusion map approximation (FPF-DM).  }
	\label{fig:SIR-FPF}
\end{figure}

Figure~\ref{fig:SIR-FPF} depicts the numerical results for the
synthetic observation data generated using the model.  Although the
results depicted in the figure are illustrative as an application of
FPF to the SIR models, additional work is necessary for its use in
prediction with real COVID-19 data.  This is because of the
following reasons:
\begin{enumerate}
	\item The observation model~\eqref{eq:obs_SIR} is not accurate.  In the real-world
	settings, one only observes a
	certain unknown (and possibly time-varying and delayed) fraction of
	the newly infected population.  This leads to fundamental issues
	with the identifiability of the transmission rate parameter
	$\beta$~\cite{roda2020difficult,wu2020substantial}.  Accurate
        estimation of $\beta$ (or the closely associated
	non-dimensional reproduction number $R_0$) is important to capture
	the initial growth of the epidemic~\cite{li2020substantial}.  
	\item The 3-state SIR dynamic model is rather simplistic. This is
	because of several reasons: (i) The model assumes homogeneous
	well-mixed population while in practice there is strong evidence of
	heterogeneities~\cite{gomes2020individual} as well as spatial network effects~\cite{pare2020modeling}; (ii) The model is based
	on the underlying assumption of Markovian transitions between the
	epidemiological states.  This assumption is contradicted by the
	experimental data on delay distributions~\cite{Olmez2020.10.03.20206250}; and (iii) Even in
	the simplistic settings of the SIR model, the transmission rate
	parameter $\beta$ is strongly time-varying.  It is affected both by
	the  individual choices (e.g., mask wearing) of the large number of
	agents as well as population-level government mandates (e.g., lockdown).   
\end{enumerate}
These difficulties notwithstanding, EnKF-based 
solutions to the COVID-19 data assimilation problem appear in~\cite{engbert2020sequential,evensen2020international}.
However, much work remains to be done on this important problem of
immense societal importance.  In the post-COVID reality, it
is not inconceivable that surveillance and monitoring of infectious
diseases such as seasonal flu will be as pervasive and common place as the
weather tracking is today.

\begin{table}
	\centering
		\caption{Simulation parameters for the application of feedback particle filter to the epidemiological example.}
	\begin{tabular}{|c|c|c|}\hline 
		parameter & notation & value \\
		\hline
		time step-size  & $\Delta t$ & 1.0 \\
		observation noise & $\sigma_W$  & $0.1$ \\
		process noise & $\sigma_B$ & $0.1$ \\
		number  of particles &  $N$& $100$ \\
		recovery rate  & $\alpha$ &  $0.1$\\
		transmission rate & $\beta$ & $0.1$ \\
		\hline
	\end{tabular}
	\label{tab:params}
\end{table}

\newpage
\processdelayedfloats 

\clearpage

\end{document}